Check for updates

# REVISED ALL-IN meta-analysis: breathing life into living systematic reviews and prospective meta-analyses

[version 2; peer review: 1 approved, 2 approved with reservations]

Previously titled: ALL-IN meta-analysis: breathing life into living systematic reviews

Judith ter Schure[1], Peter Grünwald[2,3]

[1]Epidemiology & Data Science, Amsterdam UMC, Amsterdam, North Holland, The Netherlands
[2]Machine Learning, CWI, Amsterdam, The Netherlands
[3]Mathematics, Leiden University, Leiden, The Netherlands

## Abstract

Science is justly admired as a cumulative process ("standing on the shoulders of giants"), yet scientific knowledge is typically built on a patchwork of research contributions without much coordination. This lack of efficiency has specifically been addressed in clinical research by recommendations against avoidable research waste and for living systematic reviews and prospective meta-analysis. We propose to further those recommendations with ALL-IN meta-analysis: Anytime Live and Leading INterim meta-analysis. ALL-IN provides meta-analysis based on *e*-values and anytime-valid confidence intervals that can be updated at any time—reanalyzing after each new observation while retaining type-I error and coverage guarantees, live—no need to prespecify the looks, and leading—in the decisions on whether individual studies should be initiated, stopped or expanded, the meta-analysis can be the leading source of information without losing validity to accumulation bias. The analysis design requires no information about the trial sample sizes or the number of trials eventually included. So ALL-IN meta-analysis can be applied retrospectively as well as prospectively, to evaluate the evidence once or sequentially. Because the intention of the analysis does not change the validity of the results, the results of the analysis can change the intentions ('optional stopping' and 'optional continuation' based on the results so far). On the one hand: any analysis can be turned into a living one, or even become prospective and real-time by updating with new trial data and including interim data from trials that are still ongoing—without any changes in the cut-offs for testing or the method for interval estimation. On the other hand: no stopping rule

**Open Peer Review**

**Approval Status** ? ? ✓

| | 1 | 2 | 3 |
|---|---|---|---|
| **version 2** (revision) 19 Jun 2025 | | | |
| **version 1** 19 May 2022 | ? view | ? view | ✓ view |

1. **Junfeng Wang**, Utrecht University, Utrecht, The Netherlands
2. **Ewelina Rogozinska**, University College London, London, UK
3. **Shubhendu Trivedi**, Massachusetts Institute of Technology (MIT), Cambridge, USA

Any reports and responses or comments on the article can be found at the end of the article.

needs to be enforced for the analysis to remain valid, so a prospective meta-analysis can be a bottom-up collaboration that requires no outside control over participating clinical trials. Hence ALL-IN meta-analysis breathes life into living systematic reviews and prospective meta-analyses, and offers better and simpler statistics, efficiency, collaboration and communication.

**Keywords**
Anytime, Live, Leading, Interim, Meta-analysis, Efficiency, Collaboration, Communication, Research Waste

This article is included in the Research on Research, Policy & Culture gateway.

This article is included in the Living Evidence collection.

**Corresponding author:** Judith ter Schure (j.a.terschure@amsterdamumc.nl)

**Author roles: ter Schure J**: Conceptualization, Data Curation, Formal Analysis, Methodology, Software, Visualization, Writing – Original Draft Preparation, Writing – Review & Editing; **Grünwald P**: Funding Acquisition, Investigation, Methodology, Supervision, Writing – Review & Editing

**Competing interests:** No competing interests were disclosed.

**Grant information:** This work is part of the NWO TOP-I research programme Safe Bayesian Inference assigned to Peter Grünwald, with project number 617.001.651, which is financed by the Dutch Research Counsil (Nederlandse Organisatie voor Wetenschappelijk Onderzoek; NWO).
*The funders had no role in study design, data collection and analysis, decision to publish, or preparation of the manuscript.*

**REVISED** **Amendments from Version 1**

Version 2 further illustrates the unique features of the method by extending the example ALL-IN prospective meta-analysis in the paper (ALL-IN-META-BCG-CORONA) with a reflection on the collaboration, summary of results and references to a complete replication package.

Although ALL-IN is a novel statistical approach to meta-analysis, it does provides statistical summaries for evidence synthesis that can be reported in a familiar way – *e*-values take the role of *p*-values, and anytime-valid confidence intervals the role of conventional confidence intervals. Yet due to their anytime-validity, this statistical summary can be *Live* and can be the *Leading* source of information in decisions on future research, even at *INterim* stages of clinical trials. So while the statistical approach fits within mainstream understanding of evidence synthesis as an external and retrospective summary, it can also shift that understanding towards evidence synthesis as an active/involved (advising future trials/liaising with ongoing trials) and possibly prospective tool to facilitate collaboration and increase efficiency.

In our Version 2 of the paper we have emphasized this aspect more, and moreover, stressed how the approach to prospective meta-analysis differs from other adaptive approaches to prospective analysis like the Framework for Prospective, Adaptive Meta-analysis (FAME), and Bayesian adaptive methods. The main difference is that the validity of the ALL-IN analysis does not rely on enforcing a maximum sample size or stopping rule, such that prospective collaborations can be more bottom-up rather than outsourcing decisions to a top-down meta-analysis team. This fact needed a better illustration in Section 3 on Collaboration, so we rewrote that section by extending the discussion of our BCG-CORONA example with the results from that ALL-IN meta-analysis, thus including a completely worked-out example to the paper.

**Any further responses from the reviewers can be found at the end of the article**

The scientific response to the COVID-19 (coronavirus disease 2019) pandemic constituted a major gamble. In the United States, for example, the funding program for vaccine development did not put money on a single vaccine, but on six different ones. They purposely took “multiple shots on goal” according to Larry Corey of the National Institutes of Health (NIH) COVID-19 Prevention Network in an interview with STAT (Branswell, 2021). Vaccine development is not a sure thing, and so their strategy needed to be robust enough to just “let the chips fall”. Also in the search for treatments, the scientific community had to hedge its bets. Clinical trials competed for resources and patients, and had to continuously change course when new information arrived. In contrast to vaccines, however, in most countries a strategy to find treatments was lacking. Many clinical trials suffered from “poor questions, poor study design, inefficiency of regulation and conduct, and non or poor reporting of results”: research waste (Glasziou *et al.*, 2020). We believe that more strategic thinking can benefit a future pandemic response as well as non-pandemic evidence-based medicine, as uncertainty in trial recruitment is often a given. Honest scientific bets can breathe life into the approach called *living systematic reviews*—that aims to keep the evidence record up-to-date (Elliott *et al.*, 2017) and the medical guidelines current (Akl *et al.*, 2017)—as well as *prospective meta-analysis*—that makes these ideals collaborative, prospective (Seidler *et al.*, 2019; Thomas *et al.*, 2023), and possibly ‘real-time’ (Petkova *et al.*, 2020). ‘Real-time analysis’ was promoted early in the COVID-19 pandemic and was praised by many as a “lesson learned from COVID-19” (Janiaud *et al.*, 2021; Lee *et al.*, 2022; Smith & Flaherman, 2021). We propose to achieve the necessary flexibility with *ALL-IN meta-analysis* in clinical trial design, monitoring and reporting.

ALL-IN meta-analysis stands for *Anytime Live* and *Leading INterim* meta-analysis. The *Anytime* aspect follows from the analysis with *e*-values and anytime-valid confidence intervals that can be updated at any time—reanalyzing after each new observation while retaining type-I error and coverage guarantees. This allows for *Live* adjustments to the results so far in the most flexible way possible. Because there is no need to prespecify meta-analysis looks/updates or maximum sample size, a retrospective meta-analysis can easily become a living systematic review or a prospective meta-analysis and even run in real-time based on *INterim results*. The results so far can be the *Leading* source of information: valid no matter the decisions to initiate, stop or expand individual studies, but also not enforcing any stopping rule or maximum sample size. Hence a planned prospective meta-analysis on interim results can also adjust to slow recruitment or data sharing and analyze once trials are completed (and adapt to the difficult practice of running clinical trials) as well as the ideal scenario in which a retrospective meta-analysis adjusts to a collaborative spirit in the research community and reduces research waste in a prospective real-time meta-analysis. ALL-IN meta-analysis breathes live into the possible approaches for bottom-up living collaboration that reduce avoidable research waste.

We illustrate ALL-IN meta-analysis in the setting of time-to-event data, where waiting for events is an inherent challenge of clinical trials. Combining trials early can prevent delays if studies are slow in themselves to complete the necessary number of events. ALL-IN has advantages in four categories: statistics, efficiency, collaboration and communication. We introduce all four briefly (page 5–7) before we go into more detail, but first illustrate the language of betting for single trials studying a COVID-19 vaccine.

## A single trial: the FDA COVID-19 vaccine game

On June 30th, 2020, the US Food and Drug Administration (FDA) published its guidance document on “Development and Licensure of Vaccines to Prevent COVID-19” (FDA, 2020). This set the goals for any Phase-III clinical trial betting on a protective effect of a vaccine against COVID-19. The guidance document advised on the definition of events of confirmed (symptomatic) SARS-CoV-2 infection for the trials to be counting. And in counting those, the document prescribed the two things to achieve: (1) at least a vaccine efficacy (VE) of 50% and (2) evidence against a null hypothesis of $\leq$ 30% VE (FDA, 2020, p. 14). Most COVID-19 vaccine trials randomized large numbers of participants 50:50 vaccine:placebo. So here we simplify the example by assuming that also throughout the trial the participants at risk stayed (approximately) balanced. This allows for a back-of-the-envelope calculation to reinterpret the design for the COVID-19 vaccine trials in the language of betting[1]. According to the definition of SARS-Cov-2 infections, we start counting once a participant has a confirmed infection after being fully vaccinated for at least a number of days, e.g. 7 days in the Pfizer-BioNTech trial (Polack *et al.*, 2020). This is also when a (virtual) bet could start. In the following we reinterpret the design for the COVID-19 vaccine trials in the language of betting.

Each new event carries evidence that we express by a betting score. We make a (virtual) investment on one of the two outcomes: either the next event occurs in the vaccine group or it occurs in the placebo group. If there is no effect of the vaccine whatsoever, the 50:50 risk set in our simplification ensures that the infected participant has 0.5 a chance to be vaccinated and 0.5 a chance to be a placebo. Yet, following the FDA, we do not only want to rule out an ineffective vaccine, but also reject the hypothesis that the vaccine has an effect that is too small—set as the null hypothesis of (at most) 30% VE. In that case each newly observed infection has slightly smaller chance to be a vaccinated participant. That probability to be in the vaccine group is 0.41, since each placebo group member has a 100% risk of COVID-19 and a vaccine group member has 100–30 = 70% of the risk, which is a fraction 0.41 of the total risk (70/(100 + 70)). So if the VE is too small to be of interest we expect (at least) a fraction 0.41 of COVID-19 events to occur in the vaccine group and (at most) 0.59 in placebo.

How do we bet against that and win if the vaccine has a much larger protective effect? We are betting *against* the probability 0.41 of the next COVID-19 event to occur in the vaccine group. If this probability actually is that large (the vaccine is not very protective; the null hypothesis) we do not want the game to be favorable under any strategy, just like the casino does not want any gambler to earn a salary playing the roulette wheel. On the other hand, we are betting *in favor* of a much smaller probability for the vaccine group. If this probability is smaller (the vaccine is protective; the alternative hypothesis) we do want to win money, just like a professional poker player who makes a salary out of gambling well. We use the betting scores to decide whether the vaccine is a real deal-breaker (the scores behave like the salary of a professional poker player) or whether it is not effective enough (the scores behave like anyone playing the roulette wheel). To ensure that our betting scores can show either case, we first *design* the game such that it is fair—under the null hypothesis—and then *optimize playing* the game with a strategy that is profitable—under the alternative.

**Designing a fair game under the null hypothesis** Consider gambling at the roulette table where the vaccine trial analogy is like betting on red (vaccine) or black (placebo). Betting correctly doubles your investment, betting incorrectly loses everything you risked. Assuming no house edge (no 0 or 00 on the roulette wheel) and an initial €100 you do not expect to increase your investment, since you have 0.5 a chance of doubling (2 · €100) and 0.5 a chance of losing all (0 · €100). Whether you bet everything on black or red, in expectation the betting score after one round is (0.5 · 2 + 0.5 · 0) · €100, which is the initial investment €100. To achieve the same thing betting against the 0.41:0.59 probabilities instead of 0.5:0.5, your investment needs to multiply by 2.4 (1/0.41) for vaccine and 1.7 (1/0.59) for placebo. If you bet everything on vaccine you have 0.41 chance of multiplying by 2.4 (2.4 · €100) and 0.59 chance of losing all (0 · €100) and if you bet everything on placebo you have 0.59 chance of multiplying by 1.7 (1.7 · €100) and 0.41 chance of losing all (0 · €100). The expected betting score after one round is again the initial investment for both: (0.41 · 1/0.41 + 0.59 · 0) · €100 and (0.59 · 1/0.59 + 0.41 · 0) · €100. Hence, at either the roulette table or in this FDA COVID-19 vaccine game, by design the game is fair and does not favor us. After all, if our observed infections land on the vaccine and control group with the probabilities 0.41:0.59, like a spin of the roulette wheel on black and red with 0.5:0.5, we do not expect to claim an effective vaccine.

**Optimize playing the game under the alternative hypothesis** How do we win as fast and as much as possible if our observed infections do not behave like a roulette wheel? It has been known since the work of Kelly (1956) and Breiman (1961) that the best way to increase your capital in the long run is to not bet all your (virtual) investment €100 on one of the two possible outcomes (red/vaccine or black/placebo) but to divide it based on the odds that make the game favorable to you. So our focus needs to be on the minimal VE of 50% from the FDA guidance. In the scenario of 50% VE, the probability that the next COVID-19 case is in the vaccine group is 1/3: if we set the risk of COVID-19 for a placebo group member to 100%, a vaccine group member has 100–50 = 50% of that risk, which is 1/3 of the total risk (50/(100 + 50)). Kelly (1956) and Breiman (1961) urge us to invest one-third (1/3 · €100) on observing the next infection in the vaccine group and two-thirds (2/3 · €100) on placebo.

**Likelihood ratios** If we bet this way we can rewrite our betting scores in terms of a *likelihood ratio*. We first show this for the red-black roulette game where we double what we had put at risk on either black or red if the spin of the roulette wheel outputs the color we bet on. Just like in our strategy in the FDA COVID-19 vaccine game, we put 1/3 · €100 on red and

[1] Note that in Section 1 (specifically at the end of Section 1.1) we discuss how to exactly analyze such time-to-event data using *e*-values taking into account that the risk set changes after occurrence of events and censoring. We expect such a reanalysis to be very well approximated by the simple calculations that are used for illustration in this introduction of the FDA game, at least for large and balanced trials like the one by Pfizer/BioNtech. Even for trials that are a lot smaller, this approximation is quite good. For example, the exact logrank *e*-value of 1,88 that we report for the NL trial in Table 1 can be approximated using this back-of-the-envelope calculation for the *e*-value of $\left(\frac{80/180}{100/200}\right)^{96} \cdot \left(\frac{100/180}{100/200}\right)^{110} = 1{,}33$. This is qualitatively similar to 1,88 in being much closer to 1 than to conventional thresholds like 40 or 400 (for significance level $\alpha$ = 0.025 and $\alpha$ = 0.0025). In this trial only 1496 participants were 50:50 randomized (compared to the 40 thousand in the COVID-19 vaccine trials), and 96 events observed in the treatment and 110 events in the control group (Ter Schure *et al.*, 2022); testing a 20% VE against a 0% VE.

2/3 · €100 on black, so we win the following if the ball $X$ lands on either **red** or **black**:

$$X = \textbf{red} \qquad 2\cdot\frac{1}{3}\cdot€100 = \frac{\mathcal{L}(1/3|X)}{\mathcal{L}(1/2|X)}\cdot€100$$

$$X = \textbf{black} \qquad 2\cdot\frac{2}{3}\cdot€100 = \frac{\mathcal{L}(1/3|X)}{\mathcal{L}(1/2|X)}\cdot€100$$

The Bernoulli 1/3-likelihood (1/3 | $X$) assigns likelihood 1/3 when is $X$ = **red** and 2/3 when is $X$ = **black**. So if our strategy is to invest 1/3-2/3 in roulette, our payout is our initial investment €100 multiplied by the likelihood ratio, whether $X$ is **red** or **black**. Similarly:

$$X = \textbf{vaccine} \qquad 2.4\cdot\frac{1}{3}\cdot€100 = \frac{\mathcal{L}(50\%\ \text{VE}|X)}{\mathcal{L}(30\%\ \text{VE}|X)}\cdot€100$$

$$X = \textbf{placebo} \qquad 1.7\cdot\frac{2}{3}\cdot€100 = \frac{\mathcal{L}(50\%\ \text{VE}|X)}{\mathcal{L}(30\%\ \text{VE}|X)}\cdot€100$$

The likelihood for 50% VE ((50% VE | $X$)) assigns likelihood 1/3 when is $X$ = **vaccine** and 2/3 when is $X$ = **placebo**. Similarly, the likelihood for 30% VE ((30% VE | $X$)) assigns likelihood 0.41 when is $X$ = **vaccine** and 0.59 when is $X$ = **placebo**. Hence if our strategy is to invest 1/3:2/3 in the FDA COVID-19 vaccine game, our payout is also our initial investment €100 multiplied by the likelihood ratio, whether $X$ is **vaccine** or **placebo**.

**Reinvesting** We assume now that we start with an initial (virtual) investment of €1 instead of €100, to easily assess our winnings based on the factor by which we multiply our initial investment. At the first observation we bet €0.33 on vaccine and €0.66 on placebo. After we observe the event in the placebo group we lose our €0.33 bet on vaccine and multiply our €0.66 on placebo by 1.7 to €1.13. The likelihood ratio between our 30% VE alternative hypothesis and our 50% VE null hypothesis—so (50% VE | $X$ )/(30% VE | $X$ )—is also about 1.13, so multiplying our initial investment of €1 into €1.13. On the other hand, if we observe the event in the vaccine group we lose our €0.66 bet on a placebo event and multiply our €0.33 on vaccine by 2.4 to €0.81. The likelihood ratio of a vaccine event multiplies our investment by 0.81. After each observed event we reinvest what we have left in the new bet, so multiply that with the next likelihood ratio.

**A winner** The Pfizer/BioNTech trial observed 8 cases of COVID-19 among participants assigned to receive the vaccine and 162 cases among those assigned to placebo (Polack *et al.*, 2020). This totals a betting score of $0.81^8 \cdot 1.13^{162} \cdot$ €1, which is about €118 million (note that 1.13 is really 1.13333 . . .). If someone wins that at the poker table, we have good reason to consider her a professional poker player with a favorable strategy, rather than a lucky beginner (Konnikova, 2020).

## Meta-analysis

The Pfizer/BioNTech trial included more than 43 thousand participants (Polack *et al.*, 2020), which is quite unique for a clinical trial. Usually trials are much smaller, and scientific consensus is built through systematic reviews and retrospectively combining trials in a meta-analysis. A single-trial ALL-IN analysis can be easily extended to include more than one trial, possibly in a prospective meta-analysis, and possibly on INterim data. The decision to change the intention of the original analysis by including more trials can be strategically based on the results so-far. This follows from advantages in four categories that we will first briefly introduce and then further elaborate on in this paper: statistics, efficiency, collaboration and communication.

### Statistics

Not all mRNA vaccines showed such favorable results as the Pfizer/BioNTech vaccine. In a press release CureVac AG (2021) announced that the final analysis of their clinical trial observed 83 events in the vaccinated group and 145 in placebo, so only a 43% VE (our calculations assuming a 50:50 balanced risk set ($r$ = 1 in CureVac AG (2020, p. 124))[2]). Their protocol is formulated in terms of a confidence interval for the vaccine efficacy (VE): the FDA goal is for it to exclude 30%, adjusted for two interim analyses. That adjusted confidence interval at the final analysis is [25.3%, 57.1% VE] (our calculations; normal approximation interval, based on Z α/2-statistic for the nominal level α/2 = 0.02281 (CureVac AG, 2020, Table 8)). Regrettably, this interval does not exclude 30%. When the chips fell, this trial lost.

Statistical analyses like these are essentially *all-or-nothing*, just as any other $p < \alpha$ analysis. As soon as all the $\alpha$ is spent—either on a few interims and a final analysis or just on one fixed sample size—we cannot continue the trial and perform subsequent analyses without violating the type-I error rate. This might be a reasonable price to pay in the urgency of a pandemic when multiple vaccines are competing, but it is a very inconvenient property for clinical trials in general. Usually, we do want to reanalyze a clinical trial in combination with other similar trials in a meta-analysis. Yet any $p < \alpha$ procedure is equivalent to setting a rejection region for the test statistic and checking whether the value for the statistic falls within that region. This rejection region is based on a sampling distribution that assumes the number of studies in the meta-analysis, and the number of participants within each study to be fixed in advance. Given such a fixed sample size (but also for any sequential stopping rule that sets a maximum sample size in advance, such as $\alpha$-spending), there is only one region, and your test statistic is either in it or not. If it is not, you are not allowed to redo the analyses with an increased sample size. This problem is recognized in approaches to control type-I error for living systematic reviews (Simmonds *et al.*, 2017). But also if the meta-analysis is not updated, the $\alpha$ is essentially already spent on the individual trial analyses, since the meta-analysis is an update of the trial analysis that is unscheduled and lacks type-I error control at the same level $\alpha$. If the individual study analysis would have been conclusive, the meta-analysis might never be performed, and we can recognize that we are dealing with a situation of "meta-optional-stopping". A different way to see this is by the actual sampling distribution of trials in a meta-analysis: any data-driven decision within the series—whether to accumulate more studies and when to perform the meta-analysis—changes the sampling distribution and invalidates the fixed-sample-size distribution assumed for $p < \alpha$. Hence hardly any meta-analysis has valid type-I error control, when the accumulation of trials is based on strategic decisions, in other words, those that introduce accumulation bias (Ter Schure & Grünwald, 2019).

[2] The CureVac AG (2021) press release reports a VE of 48%, so uses a different $r$ (ratio of follow-up time in the two groups). In such large trials $r$ can often be assumed to stay close to 1, so we set it to 1 to make all calculations simpler. All our calculations are available as R code in the software availability statement. (Ter Schure, 2025, https://doi.org/10.17605/OSF.IO/D9JNY).

ALL-IN meta-analysis is not *all-or-nothing* and can combine all available studies, whatever the intentions of earlier analyses. In fact, it allows any number of new studies or participants to be included without ever spending all $\alpha$. In terms of gambling, we can keep betting our virtual investment because we never lose everything. The CureVac AG (2021) results, for example, would have accumulated a betting score of $0.81^{83} \cdot 1.13^{145}$. €1 = €1.84. In the setting of a pandemic with competing vaccines in large trials, this research was not continued. But in other settings, promising, but inconclusive results appear often in smaller trials that should spur new research. The analysis can preserve the evidence to reinvest in the next trial, such that we can continue to observe evidence and express it by betting on additional observations in a new trial. An ALL-IN meta-analysis can always continue testing the null hypothesis—with type-I error control—and estimating the confidence interval—with coverage guarantees. Importantly, for these tests and intervals the procedures are exactly the same no matter what decisions—so-called stopping rules, or accumulation bias processes (Ter Schure & Grünwald, 2019)—are at play. This is not the case for any other approach to type-I error control in living systematic reviews.

**Bottom-up living meta-analysis** The lack of restrictions on sample size is a unique feature of ALL-IN meta-analysis that sets it apart from other statistical approaches to living systematic reviews (Simmonds *et al.*, 2017), 'adaptive' (Tierney *et al.*, 2021) or 'real-time' (Petkova *et al.*, 2020) prospective meta-analysis and 'metatrials' (van Haren *et al.*, 2021). The only approach that is similar was the proposal of the law of iterated logarithm in meta-analysis (Hu *et al.*, 2007; Lan *et al.*, 2003) and is based on early work by Robbins and colleagues (Robbins, 1970) that is closely related to the methods proposed here (see e.g. Wang & Ramdas (2025) for a comparison with early work). (Our representation based on betting scores/likelihood ratios/*e*-values has advantages in the ease of communication and its rapidly improving field of statistical methods). Other approaches actually in use require a maximum sample size or a maximum number of studies to guarantee type-I error control for all updates of the meta-analysis. The crucial difference with ALL-IN is that for those methods to be valid, participating in a prospective meta-analysis requires outside control over a trial's data collection. Outside control is needed in the case of the 'Framework for prospective adaptive meta-analysis' (FAME) (Tierney *et al.*, 2021) where a single sample size is set comprised of data from all ongoing trials, possibly at an interim of some of them, and single-analysis statistics is used. The consequence is that the intervals and *p*-values reported are only valid if never updated, so the analysis assumes that data collection stops after that point. No new analyses are valid after that maximum sample size, since all alpha is spent and possibly accumulation bias is introduced in a re-analysis (Ter Schure & Grünwald, 2019). Similarly, outside control is needed in the operationalization of group-sequential or alpha-spending methods that need the information size relative to the final analysis (Simmonds *et al.*, 2017), and therefore also need to enforce such maximum sample size for the reported results to be valid. Finally, even in in simulation-calibrated Bayesian approaches (Goldfeld *et al.*, 2021; van Haren *et al.*, 2021) restrictions on interims (often) and maximum sample size (always) are imposed for the simulations to terminate. So the frequentist operating characteristics (type-I error, coverage of intervals) of the analysis are not known if trials wish to continue data collection after that maximum sample size and analyze again, or in other ways deviate from the simulated scenarios. Hence all available methods in living systematic reviews and prospective meta-analysis are by design quite top-down, or even explicitly described as "using preestablished stopping rules for safety, efficacy, futility, and harm" (Petkova *et al.*, 2020).

The fact that ALL-IN does not require such outside control for the analysis to be valid not only simplifies the statistics, it also simplifies collaboration as a bottom-up instead of a top-down process. We will illustrate this further in our worked-out example in Section 3 on Collaboration.

## Efficiency

Lack of efficiency has been addressed in clinical research in many ways. Not only in the proposal of living systematic reviews (Elliott *et al.*, 2017), but also in encouragements to present new studies in the context of existing evidence (Young & Horton, 2005), in advice to design new trials based on systematic reviews and meta-analysis (Chalmers & Lau, 1993; Goudie *et al.*, 2010; Lau *et al.*, 1995; Lund *et al.*, 2016; Sutton *et al.*, 2007) and in pleading to prevent the "scandal" of wasteful research into clinical questions that are already answered or not of primary importance (Altman, 1994; Chalmers & Glasziou, 2009; Glasziou & Chalmers, 2018; Glasziou *et al.*, 2020; Ioannidis *et al.*, 2014, "research waste"). These calls have not been completely ignored, since clinical research has seen an increase in efficiency—e.g. in platform trials or prospective meta-analysis whenever collaboration is deemed possible prospectively. Nevertheless, most clinical trial data is synthesized retrospectively, and still deserves all of the above recommendations. ALL-IN meta-analysis enables these data-driven decisions that can make science more efficient; any retrospective meta-analysis can become prospective. New studies can be easily informed by the synthesis of all data so far such that exactly the right number of participants are randomized to answer a research question, no more and no less. Moreover, an ALL-IN meta-analysis can give an account of the evidence at any time and therefore facilitate prioritizing new studies, if more than one line of research needs additional data, but not all can be funded.

## Collaboration

Any ALL-IN meta-analysis can easily be turned into a prospective meta-analysis and possibly a *live* or real-time meta-analysis, since it does not matter how many studies will eventually be combined or which study will contribute most data. Whether it is based on summary statistics (Godolphin *et al.*, 2022; Tierney *et al.*, 2021) or on individual participant data (IPD) (Polanin & Williams, 2016), involvement in the same prospective meta-analysis facilitates discussion between those running trials in the same line of research; especially if the line of research can be concluded early. Collaboration in a prospective meta-analysis (PMA) has many advantages, such as reducing research waste and bias (by specifying inclusion criteria and analysis plans in advance) and harmonizing outcome measures (only possible in a close collaboration with trials). Live or real-time PMA can increase the involvement of all trials in such a collaboration, when trials are motivated by the promise of a possible early conclusion (Seidler *et al.*, 2019 and Thomas *et al.*, 2023).

A prospective meta-analysis benefits from homogeneity. With too much heterogeneity, it can be very disheartening to update a random-effects meta-analysis, since many trials are needed to precisely estimate the between trial variation and shrink the interval (Jackson & Turner, 2017; Kulinskaya & Wood, 2014; Sutton *et al.*, 2007). Close collaboration might prevent unnecessary heterogeneity, if trial investigators are involved in the selection of trials in the meta-analysis; especially if they can advise on the design and conduct of new trials and align inclusion criteria and endpoint definitions. A fixed-effects meta-analysis can conclude the research effort early. Sufficient homogeneity may be possible in close collaboration.

## Communication

**The language of betting** The interpretation of evidence in terms of a betting score might help to communicate the uncertainty in statistical results. As Shafer (2021) puts it: “When statistical tests and conclusions are framed as bets, everyone understands their limitations. Great success in betting against probabilities may be the best evidence we can have that the probabilities are wrong, but everyone understands that such success may be mere luck.” Thinking in terms of bets also helps to understand when statistical analyses can be *anytime-valid*. If they are of the *all-or-nothing* kind, but reanalyzed in a meta-analysis, they are gambling while broke. (This intuition can be made mathematically precise; see the description of Neyman-Pearson testing in terms of betting Shafer (2021) and Grünwald *et al.* (2024a)). Yet if we add new studies to an ALL-IN meta-analysis, we are reinvesting the betting score that we saved from earlier studies, to evaluate whether the strategy in those earlier studies continues to succeed. Just like when reinvesting your profits in a casino from one slot machine into another, the notion of winning stays the same. Our evidence against the hypothesis of a *fair* casino does not change when we alternate slot machines. It does not change if we use the score so far to decide on alternating them or to decide when to cash out. If the slot machines are fair, any strategy of playing them is not expected to make money, and our notion of type-I error control holds under any dependency on past results (stopping rules or accumulation bias processes). This is the intuition that explains why it is statistically possible for the intensions of the analysis to not change its validity, such that the analysis results can change the intentions. Any meta-analysis can become a living systematic review, possibly prospective and possibly real-time.

**Other communication** Those uncomfortable with the language of betting can also easily resort to any of three more familiar notions of statistical communication. Firstly, the likelihood ratios/betting scores and their generalizations, so-called *e-values* (Grünwald *et al.*, 2024a; Ramdas & Wang (2024); Vovk & Wang, 2021), can be interpreted as conservative *p*-values by taking their inverse. If we denote any betting score or *e*-value by € (e.g. € = 1.84 for the CureVac trial data), then $p < 1/€$ is a conservative *p*-value (e.g. p = 1/1.84 = 0.54 for the CureVac trial data). If we communicate the *p*-value $p = 1/€$ anyone can test by comparing $p < \alpha$ but with the addition that this conservative *p*-value is anytime valid[3] and so $p < \alpha$ can never spend all $\alpha$ (it is never an *all-or-nothing* test). Secondly, the likelihood ratios have their own notion of evidence in the likelihood paradigm (Royall, 1997). Just as well as stating that the Pfizer/BioNTech trial (Polack *et al.*, 2020) multiplied €1 to almost €118 million and the CureVac AG (2021) trial multiplied €1 to €1.84, we can state that their data was almost 118 million times and 1.84 times more likely if we assume the FDA’s goal of 50% VE in comparison to assuming only 30% VE. For Pfizer, that sounds very good, for CureVac, not so much, and so these numbers have an interpretation of their own without imposing any $\alpha$-level. Thirdly, likelihood ratios can be accepted by the Bayesian paradigm, as Bayes factors, and possibly combined with prior odds. Grünwald *et al.* (2024a) and Grünwald (2021) show that betting scores/*e*-values and Bayes factors are closely related, although not all Bayes factors are betting scores/*e*-values. The bottom-line for communication purposes is that the reporting by ALL-IN meta-analysis can be interpreted in many ways—*p*-values, likelihood ratios, Bayes factors—but regardless of the interpretation provide fully frequentist type-I error control for tests and coverage for anytime-valid confidence intervals.

The remainder of this paper discusses the four categories of advantages in more detail: *Statistics* in Section 1, *Efficiency* in Section 2, *Collaboration* in Section 3 and *Communication* in Section 4. We use the COVID-19 vaccine trials as running examples, based on the FDA COVID-19 vaccine game described already, but also in terms of the *e-value logrank test* (Ter Schure *et al.*, 2024). In section Section 3 on Collaboration we discuss an example ALL-IN meta-analysis that used this *e*-value logrank test to study whether the Bacillus Calmette–Guérin (BCG) vaccine, originally developed to protect against tuberculosis and named after its inventors, could protect against COVID-19 (ter Schure *et al.*, 2022). In the concluding section we will provide some broader context, with an overview of all the methods already developed—*e*-values, *safe* tests (Grünwald *et al.*, 2024a) and *anytime-valid* confidence intervals—methods already available in software—notably safestats R package (Turner *et al.*, 2022)—and future work. R code for all calculations, simulations and plots is available through the software availability statement and Ter Schure, 2025, https://doi.org/10.17605/OSF.IO/U6WTP.

[3] Such conservative *p*-values cannot be pictured as the tails of a sampling distribution since such a picture needs a sample size. Also, for conventional *p*-values, 1/p is not an *e*-value. The Introduction chapter and the appendix to Chapter 1 in the Ph.D. dissertation Ter Schure (2022b) give more details.

## 1 Statistics

The language of betting comes with the intuition that winning a large betting score has a small probability if the null hypothesis is generating our observations (e.g. the roulette wheel is fair). We will make this intuition precise and show how to control the type-I error by bounding this probability by Markov's inequality and Ville's inequality. Crucial here is that the betting score underlying our test is an *e-value*. The language of betting also comes with the intuition that when playing a game that is favorable to us in principle, we can use strategies of different quality: even among all strategies under which we expect to get richer, some of them can be expected to earn us much more than others. We will relate the more well-known notion of power to such a different notion of *optimality*. In the following we discuss both *e-values* and *optimality* first for a single trial (in the FDA COVID-19 vaccine game and more generally) and then for ALL-IN meta-analysis. We conclude by a generalization of optimal *e*-value tests to anytime-valid confidence intervals.

### 1.1 Under the null: *e*-values in a single trial

To make the FDA COVID-19 vaccine game fair we imposed a multiplication by 2.4 (or 170/70) if we observe the event in the vaccine group and 1.7 (or 170/100) if we observe it in the placebo group. This multiplication has expectation 1 (or smaller) if we assume the null hypothesis of a vaccine with negligible VE of 30% (or smaller). In case of 30%, we have probability 0.41 (or 70/170) to observe a vaccine event and probability 0.59 (or 100/170) to observe placebo, so in expectation we multiply our investment by 1. For example putting 1/3 on vaccine and 2/3 on placebo: 1/3· 70/170 · 170/70 + 2/3· 100/170 ·170/100 = 1. No matter how we invest in the two outcomes, (e.g. try putting 1/2 on vaccine and 1/2 on placebo, or something different) in expectation under the null we multiply the initial investment by 1. This means that our betting score is an *e*-value, since by definition an *e*-value is the outcome of a nonnegative random variable with expectation (at most) 1 under the null hypothesis (Grünwald *et al.*, 2024a; Ramdas & Wang, 2024; Vovk & Wang, 2021).

Our betting score could also be rewritten as a likelihood ratio, so the expectation of the likelihood ratio ($\mathcal{L}$(50% VE | $X$)/$\mathcal{L}$(30% VE | $X$)) is 1 as well. We hence-forth write the likelihood ratio after $n$ rounds of betting (or after observing $n$ events) as $\mathbf{LR}^{(n)}$, with for the FDA COVID-19 vaccine game

$$\mathbf{LR}^{(n)} = \prod_{i=1}^{n} \frac{\mathcal{L}(50\%\ \mathrm{VE}\,|\,X_i)}{\mathcal{L}(30\%\ \mathrm{VE}\,|\,X_i)}. \tag{1}$$

Using its expectation of 1, Markov's inequality bounds the probability of observing a large multiplication of our investment (a large likelihood ratio) by $\alpha$ after $n = 170$ rounds as follows:

$$\mathbf{P}_{30\%\ \mathrm{VE}}\left[\mathbf{LR}^{(170)} \geq 1/\alpha\right] \leq \frac{\mathbf{E}_{30\%\ \mathrm{VE}}\left[\mathbf{LR}^{(170)}\right]}{1/\alpha} = \frac{1}{1/\alpha} = \alpha.$$

Figure 1 shows at the right side the histogram of betting scores in the FDA COVID-19 vaccine game after 170 events when we simulate events under the null hypothesis, with probability 0.41 to occur in the vaccine group, corresponding to 30% VE. A line is shown at 40, and indeed no more than $\alpha$ = 1/40 = 2.5% of the scores seem to be larger than that threshold. In fact, in these 1000 runs of simulation only 0.3% of the runs have a betting score larger than 40; Markov's inequality is a loose bound. We also have a stronger result because we obtained our betting score over events by multiplying the score of the rounds (see (1), corresponding to reinvesting our

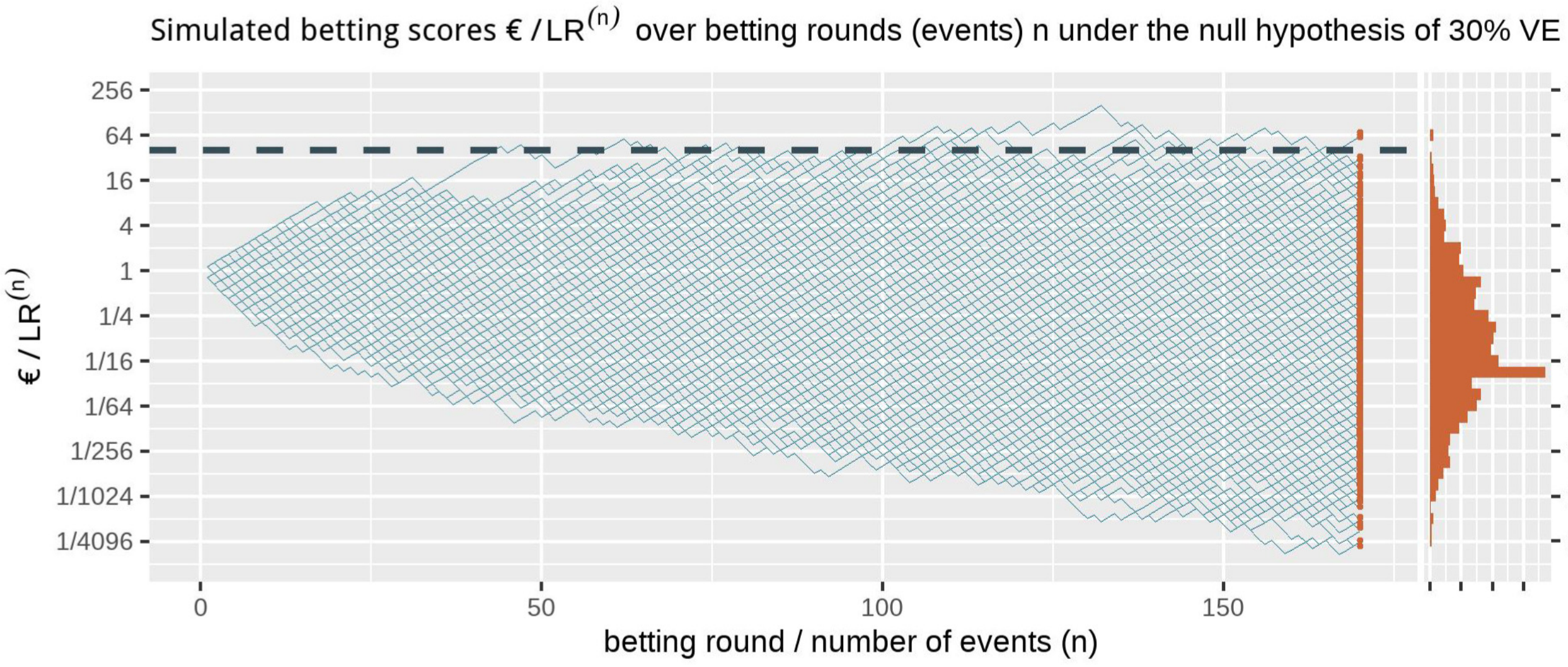

**Figure 1. 1000 simulated betting scores in the FDA COVID-19 vaccine game over betting rounds *n* assuming a probability of 0.41 (70/170) for each event to occur in the vaccine group (the null hypothesis of 30% VE) and optimal Kelly betting for 50% VE.** The dashed line is the threshold 1/$\alpha$ = 40 one-sided. The histogram at the right shows the betting score/$LR^{(170)}$ after 170 events. Note that the expectation of 1 of the scores is not the mode of its distribution nor its median and that the vertical axis is on a logarithmic scale.

winnings), called Ville's inequality. We get the following from Ville (1939):

$$\mathbf{P}_{30\%\ \mathrm{VE}}\left[\mathbf{LR}^{(n)} \geq 1/\alpha \quad \text{for some } n\right] \leq \alpha.$$

Ville's inequality is also illustrated in Figure 1: if we take the sequence of rounds into account, still only a few out of the 1000 simulations *ever* reach a betting score larger than 40. In fact, in these 1000 runs of simulation only 1.1% of the runs have a betting score that is larger at any round in the game, such that our type-I error is controlled at $\alpha = 2.5\%$ at any time. Moreover, this type-I error control is not tied to this maximum number of 170 events, but continues to hold with an unlimited horizon. Making a large profit in such a fair game casts doubt on the null hypothesis and is captured by a likelihood ratio that grows away from 1: a large betting profit is obtained if the null likelihood is performing worse than alternative.

**When trials can be summarized as bets** Before they can be combined in a meta-analysis, individual trials are often characterized by the summary statistics from trial publications. Conventional meta-analysis combines these statistics (e.g. mean differences and standard deviations) in a $Z$-statistic (Borenstein *et al.*, 2009). Unlike the vaccine/placebo outcomes that we have seen so far, such a $Z$-statistic has a continuous density and cannot be summarized by separately dealing with all possible outcomes. Fortunately, Shafer (2021) shows that any likelihood ratio of distributions can be viewed as a betting score in a game with initial investment €1. This is possible because likelihood ratios have expectation 1 in general if we assume the null hypothesis in the denominator of the ratio to generate the data. For a $Z$-statistic we have two normal distributions with variance 1, one with mean $\mu_0$ under the null hypothesis, and one with $\mu_1$ under the alternative. If the data is generated by the null model, the expectation of the likelihood ratio is

$$\mathbf{E}_{Z\sim\phi_{\mu_0}}\left[\frac{\phi_{\mu_1}(Z)}{\phi_{\mu_0}(Z)}\right] = \int_z \phi_{\mu_0}(z)\frac{\phi_{\mu_1}(z)}{\phi_{\mu_0}(z)}\mathrm{d}z = \int_z \phi_{\mu_1}(z)\mathrm{d}z = 1, \qquad (2)$$

since $\phi_{\mu 1}(z)$ is a probability density that integrates to 1. This means that any such likelihood ratio for a $Z$-statistic is an $e$-value and can be used to construct tests by betting.

Not all summary statistics can be assumed to form a $Z$-statistic with a normal distribution. Fortunately for the logrank statistic this is reasonable (Ter Schure *et al.*, 2024) if studies are large and the effect size not too extreme (hazard ratios not too far away from 1). We will use the logrank $Z$-statistic as a running example for meta-analysis on summary statistics. For an IPD meta-analysis (on individual participant data), however, we recommend to use the exact $e$-value logrank test from Ter Schure *et al.* (2024) that is valid regardless of the randomization (e.g. 1:1 balanced or 1:2 unbalanced), the number of participants at risk, the number of events or the size of the effect.

## 1.2 Under the null: *e*-values in a (live) meta-analysis

Assume we want to perform a meta-analysis and we collect a $Z$-statistic $Z_i$ from each trial $i$, e.g. a logrank statistic. Before observing $Z_i$ we construct an honest bet $\mathbf{LR}_i = \phi_{\mu_1}(Z_i)/\phi_{\mu_0}(Z_i)$ for each trial that is an $e$-value and thus has type-I error control under the null hypothesis $\phi_{\mu 0}$—for a default logrank statistic this is always $\mu_0 = 0$ corresponding to hazard ratio of 1. If we think of the betting score from the first study and invest it in the second study, we are in fact multiplying likelihood ratios. We need to have a notion of time $t$, such that at each time we know the number of studies $k\langle t\rangle$ so far and the number of observations $n_i\langle t\rangle$ in each study $i$. Note that for the logrank statistic, and time-to-event in general, the number of observations or sample size $n$ is the number of events. If we assume that all studies are completed at time $t$ with $n_1, n_2, \ldots, n_k$ events summarized by logrank $Z$-statistics $z_1^{(n_1)}, z_2^{(n_2)}, \ldots, z_k^{(n_k)}$ we can construct our ALL-IN bet as follows:

$$\mathbf{LR}_{\mathrm{META}}^{\langle t\rangle} = \prod_{i=1}^{k\langle t\rangle}\mathbf{LR}_i^{(n_i)} = \prod_{i=1}^{k\langle t\rangle}\frac{\phi_{\mu_1\sqrt{n_i}}(z_i^{(n_i)})}{\phi_0(z_i^{(n_i)})}. \qquad (3)$$

**The global null hypothesis** Each trial bet is testing the same null hypothesis $\mu_0 = 0$ in (3), such that the ALL-IN meta-analysis bet tests a *global null hypothesis* of no effect (0% VE, hazard ratio 1) in all trials. Such a global null hypothesis can be rejected with a contribution from each trial, but also in case only one trial observes a large score betting against the hypothesis and no other trial observes a very small betting score that loses those winnings again. After all, the global null (null in each trial) is rejected as soon as the null is rejected in one of the trials.

**Meta-analysis on interim data** We can generalize this ALL-IN meta-analysis bet of completed trials to bets on interim data by assuming that we only have an interim logrank $Z$-statistic $z_1\langle t\rangle, z_2\langle t\rangle, \ldots, z_k\langle t\rangle$ for the $n_1\langle t\rangle, n_2\langle t\rangle, \ldots, n_k\langle t\rangle$ events observed so far at time $t$; $k\langle t\rangle$ still represents the number of studies so far at time $t$, but now these studies are not (all) completed. We construct our ALL-IN bet in a similar way:

$$\mathbf{LR}_{\mathrm{META}}^{\langle t\rangle} = \prod_{i=1}^{k\langle t\rangle}\mathbf{LR}_i^{(n_i\langle t\rangle)} = \prod_{i=1}^{k\langle t\rangle}\frac{\phi_{\mu_1\sqrt{n_i\langle t\rangle}}(z_i\langle t\rangle)}{\phi_0(z_i\langle t\rangle)}. \qquad (4)$$

From the perspective of Ville's inequality, the analysis on completed trials and the one on interim data are indistinguishable. The only thing that matters is that we include all the data we have so far at time $t$, such that we have type-I error control

$$\mathbf{P}_0\left[\mathbf{LR}_{\mathrm{META}}^{\langle t\rangle} \geq 1/\alpha \quad \text{for some } t\right] \leq \alpha, \qquad (5)$$

for the global null hypothesis probability $\mathbf{P}_0$ with an unlimited horizon over time $t$. This is why any ALL-IN meta-analysis can become prospective and live or real-time on INterim data. The reported analysis is the same, no matter the intentions of the analysis.

## 1.3 Under the alternative: optimality in a single trial

A power analysis sets a very specific goal for a trial, usually to detect an effect of minimal clinical importance (MCID, minimal-clinically important difference) or smallest effect size of interest (SESOI). This is the effect we would not like to miss if it were there, although we hope that the real effect is larger. We nevertheless use this effect size of minimal clinical importance to decide on the sample size of the trial, otherwise we risk a futile trial. The FDA was clear on what this minimal effect should be for the COVID-19 vaccine trials: a VE of 50% (FDA, 2020). This is the effect we used to bet in the FDA COVID-19 vaccine game.

Our strategy in the FDA COVID-19 vaccine game, however, was not trying to achieve optimal power. If we compare the *all-or-nothing* confidence interval for CureVac AG (2021) from the introduction—the final analysis on 83+145 events—we notice that this confidence interval [25.3%, 57.1% VE] is smaller than the final anytime valid confidence interval we show in Figure 3 in Section 1.5, which is [17.1%, 61.6% VE][4]. The difference is that the former one is optimized to have spent all $\alpha$ at the final analysis, while the latter one is optimized to continue data collection. Power is the probability of finding the desired result using the specified analysis at a sample size or stopping rule. So for an analysis that is intended to have unlimited horizon, power is not a well-defined concept. Instead the literature on *e*-values (Grünwald *et al.*, 2024a) optimize the expected rate at which the evidence grows for each new data point, not at a specific sample size (Grünwald *et al.* (2024a) called *growth-rate optimality (GRO)* in Grünwald *et al.* (2024a), *log-optimality* in Ramdas & Wang (2024), and the unique *e*-value that is optimal in this way the *numeraire* (Larsson *et al.*, 2024)). The worst case here is the 50% VE for a one-sided alternative hypothesis $H_1 = \{P_{\text{VE}} : 50\% \leq \text{VE} \leq 100\%\}$. We optimized the FDA bet in the introduction by putting this 50% VE in the alternative likelihood. This can be rewritten in terms of a likelihood ratio for the logrank statistic $Z$ as follows:

$$\mathbf{LR}^{(n)} = \prod_{i=1}^{n} \frac{\mathcal{L}(50\%\ \text{VE} \mid X_i)}{\mathcal{L}(30\%\ \text{VE} \mid X_i)} = \frac{\mathcal{L}(50\%\ \text{VE} \mid X_1, \ldots, X_n)}{\mathcal{L}(30\%\ \text{VE} \mid X_1, \ldots, X_n)} \approx \frac{\phi_{\mu_{\min}\sqrt{n}}(Z^{(n)})}{\phi_{\mu_0\sqrt{n}}(Z^{(n)})}, \tag{6}$$

with $\mu_{\min} = 1/2 \log(0.5)$ and $\mu_0 = 1/2 \log(0.7)$ with 0.5 and 0.7 the hazard ratios corresponding to VE of 50% and 30% respectively (see Ter Schure *et al.* (2024)). So our one-sided alternative hypothesis for the logrank $Z$-statistic is a $Z$-distribution with a mean representing an effect that is at least $\mu_{\min}$ ('at least' here means more negative, since for a hazard reduction $\mu_{\min}$ is negative):

$$H_1 = \{\phi_{\mu_1} : \mu_1 \leq \mu_{\min}\}$$

(since positive VE corresponds to a negative $\mu$). Our choice of the parameter of the alternative likelihood $\mu_{\min}$ follows directly from the minimal effect set by the FDA. Kelly (1956) already showed that this way of betting optimizes the way our betting score grows if the true VE is 50% (our worst-case scenario). Breiman (1961) showed that this approach also minimizes the expected number of events we need to reach a given betting score set in advance (e.g. €$1/\alpha$), for which some intuition is given in Figure 2. Grünwald *et al.* (2024a), Shafer (2021) and the appendix to Ter Schure *et al.* (2024) give various other reasons why this is the best way to bet, relating it to data compression, information theory, Neyman-Pearson testing, Gibb's inequality, and Wald's identity. The most crucial property for the purposes of ALL-IN meta-analysis is that the alternative likelihood puts some money on each possible outcome, such that no matter what outcome we observe, we keep some of the money we risk: our betting score can become small but not 0. This contrasts the approach with a classic $p < \alpha$ test that essentially puts all money on the rejection region, such that if the outcome is not in it, we lose all and cannot continue betting. A thorough interpretation of Neyman-Pearson testing and $p$-values in terms of betting is given by both Grünwald *et al.* (2024a) and Shafer (2021).

[4] Note that we are comparing a $Z_{\alpha/2}$-confidence interval for $\alpha/2$ = 0.02281 with $\alpha/2$ = 0.025, so the wider interval cannot be attributed to the level of $\alpha$, because for the same method of interval construction we get wider intervals for smaller $\alpha$.

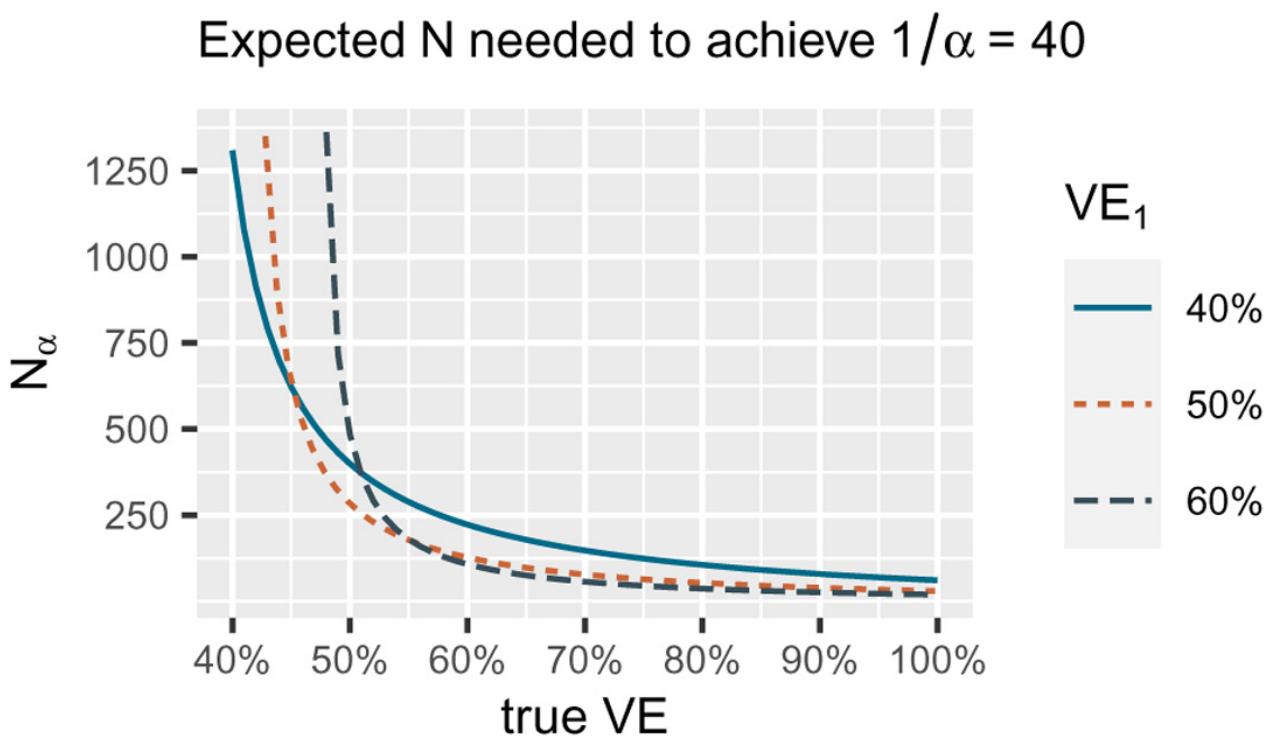

**Figure 2. $N_\alpha$ is the expected number of events needed to reach a betting score of $1/\alpha$ = 40 for $\alpha$ = 0.025 if we bet according to $\text{VE}_1$ indicated by the three different lines, with bets each of the form** $\prod_{i=1}^{N} \frac{\mathcal{L}(\text{VE}_1 \mid X_i)}{\mathcal{L}(30\%\text{VE} \mid X_i)}$. The number of events we need decreases if the true Vaccine efficacy (VE) underlying the data increases (the true difference in risk between vaccine and control is larger). The smallest number of events for a true VE of 40% is reached by betting $\text{VE}_1$ of 40% (blue solid line), the smallest number of events for a true VE of 50% by betting $\text{VE}_1$ of 50% (orange dotted line) and the smallest number for true VE of 60% by betting $\text{VE}_1$ of 60% (grey dashed line). Note that for the alternative in the FDA COVID-19 vaccine game $H_1 = \{P_{\text{VE}} : 50\% \leq \text{VE} \leq 100\%\}$ we are only interested in playing the game well if the true VE is 50% or larger. Since for larger true VE, taking $\text{VE}_1$ = 50% performs quite well, our strategy is to optimize for the worst case of 50% VE itself and use the bet with $\text{VE}_1$ = 50% in the FDA COVID-19 vaccine game. This figure uses Wald's identity. Specific calculations are available in R code on OSF linked in the Software availability section.

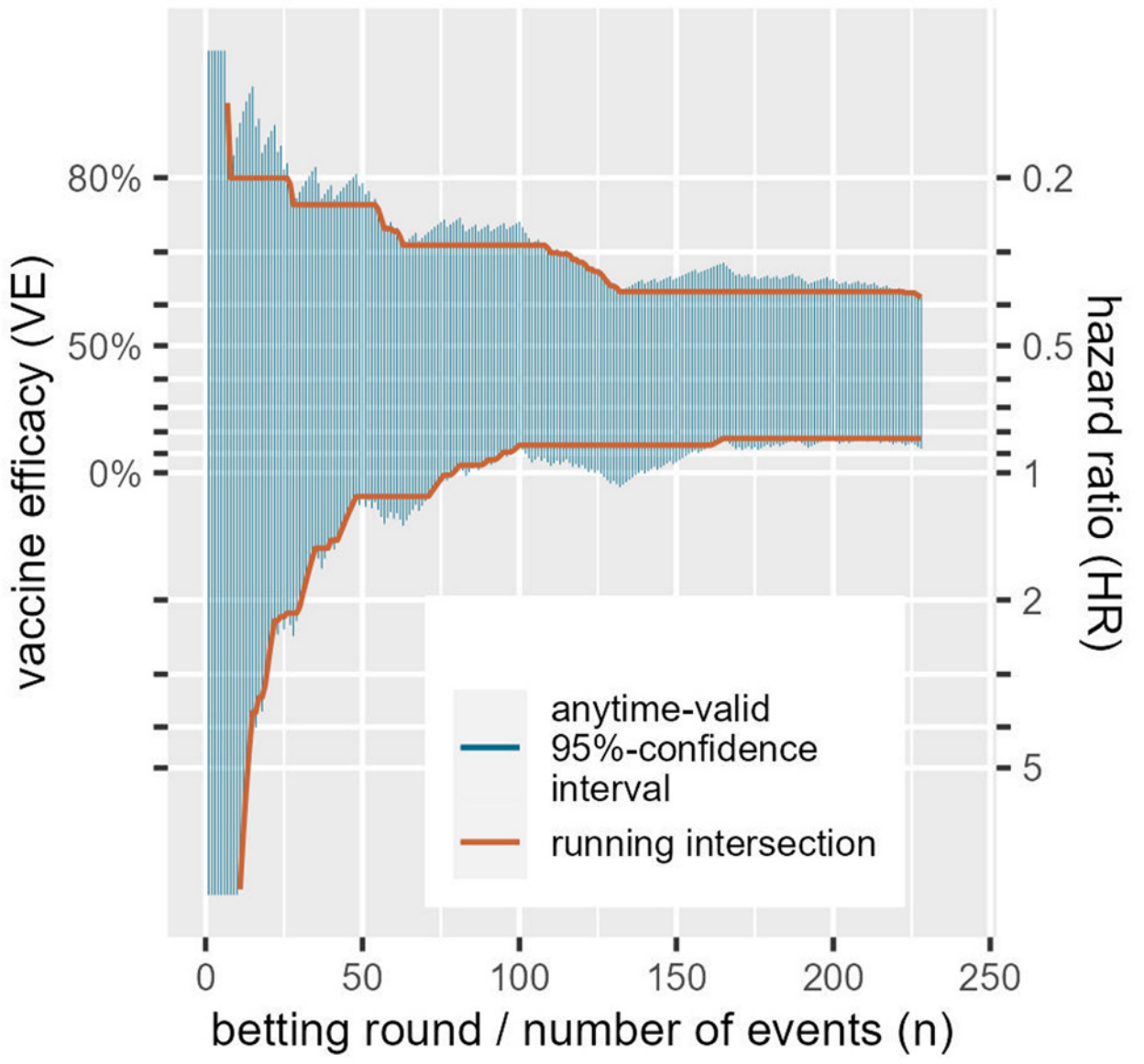

**Figure 3. Anytime-valid 95%-confidence intervals for a random ordering of the 83 events in the vaccine group and 145 events in placebo from the CureVac AG (2021) trial.** Note that the vertical axis is on a logarithmic scale and that the hazard ratio scale on the right is flipped: smaller values at the top mean smaller risk in the CureVac vaccinated group and therefore larger vaccine efficacy.

### 1.4 Under the alternative: optimality in a meta-analysis

ALL-IN meta-analysis allows for a retrospective meta-analysis that is bottom-up. The betting score that we accumulate by reinvesting from one trial into the other (which is multiplying betting scores) has an interpretation without enforcing a common design or stopping rule on all included trials. This is especially important if trials have their own stopping rules, or if accumulation processes are at play that influence the existence of trials based on earlier (trial) results in the same meta-analysis. While a meta-analysis can be bottom-up and each have its own design and effect of minimal interest, it can be advisable to agree on a $\mu_{\min}$ for the meta-analysis. However, the meta-analysis betting score can also allow each trial $i$ to have its own alternative likelihood with parameter $\mu_{\min(i)}$. Then the following multiplication of those betting scores is still a valid meta score with type-I guarantees:

$$\mathbf{LR}_{\text{META}}^{\langle t\rangle} = \prod_{i=1}^{k\langle t\rangle} \frac{\phi_{\mu_{\min(i)}\sqrt{n_i}}(z_i^{(n_i)})}{\phi_0(z_i^{(n_i)})}. \tag{7}$$

As long as $\phi_{\mu_{\min(i)}\sqrt{n_i}}$ is a probability density that integrates to 1, we have that each likelihood ratio integrates to 1 under the global null hypothesis, such that (5) holds. This means that trials can also learn their parameter $\mu_{\min(i)}$ from already completed trials. Sometimes trials are not powered to detect an effect of minimal interest, but an effect that is plausibly true based on earlier research. Kulinskaya *et al.* (2016) shows that such use of existing studies to power new trials can actually bias conventional meta-analysis since it introduces yet another dependency between sample size and results that is unaccounted for in any analysis that assumes a fixed sample size. For ALL-IN meta-analysis this is no problem at all, and trials can learn from each other as long as the parameter $\mu_{\min(i)}$ is fixed before seeing new data that is evaluated using that parameter in (7). In ter Schure *et al.* (2024) we discuss the advantages of even learning the parameter within one trial using prequential plugins or Bayesian updating. In a game like the FDA COVID-19 vaccine game with a clear goal, this is inferior to GRO/log-optimality, but in other situations it could be preferred.

### 1.5 Anytime-valid confidence intervals

The CureVac AG (2021) trial reached their final interim analysis but was not able to reject the null hypothesis of 30% VE. The trial had also been optimistically powered for 60% instead of 50% VE (CureVac AG, 2020). If a trial is underpowered but still has a large number of participants in follow-up, there is good reason to continue the trial, or combine the trial with results from a new trial in a meta-analysis. However, with a total of 227 events this trial was not underpowered to reject the null hypothesis with an effect in the same ballpark as the Pfizer/BioNTech trial that reported 95% VE (Polack *et al.*, 2020). In such a case it is very interesting to zoom in on the estimate for the effect, instead of its test.

A standard confidence interval can be seen as an inversion of a hypothesis test: if the null falls outside a two-sided 95%-confidence interval it can be rejected with a one-sided type-I error level of $\alpha/2 = 0.025$. In general, the interval excludes all the values for the parameter that can be rejected when representing the null hypothesis. Similarly, in our context, an anytime-valid confidence interval excludes all values of the parameter that can be rejected by the $e$-value test that corresponds to the betting strategy at hand. So the interval is essentially tracking a whole range of bets, each against a different null hypothesis. Figure 3 gives a sequence of anytime-valid 95%-confidence intervals for a random ordering of the CureVac AG (2021) data, one for each new observed event or betting round. It shows that the more events we observe, the more parameter values (values for hazard ratios, or their corresponding VEs) we can exclude from the interval. Because these intervals are valid at any time, once we can exclude a value, we never have to include it again. So we also show a sequence of intervals that is the running intersection of all the previous intervals. This of course crucially depends on the ordering, so the one shown for the CureVac AG (2021) data is just an example, since the ordering is randomly chosen. Since these intervals are anytime valid, it is possible to further shrink the intervals by continuing follow-up and observing more events. The coverage of an anytime-valid confidence interval—like an $e$-value test—has an unlimited horizon.

An ALL-IN meta-analysis confidence interval that is based on a running intersection is of course only possible in an IPD meta-analysis, and cannot be based on summary statistics. The confidence interval shown in Figure 3 is based on the logrank $Z$-statistic (by repeatedly calculating it after each event), which can also be a summary statistic to achieve a single interval that is anytime-valid. The interval follows from the likelihood ratio of normal densities from (6) and follows a general recipe for constructing anytime-valid confidence intervals from Howard *et al.* (2021) where the hazard ratio is obtained by maximum-likelihood. The same approach can be used to obtain an ALL-IN meta-analysis confidence interval by weighing the estimates by inverse-variance (the fixed-effects approach) that Peto (1987) calls a *typical* hazard ratio. We illustrate this approach in our example ALL-IN meta-analysis in Section 3 on Collaboration (for technical details see the *Statistical Appendix* of ter Schure *et al.*, 2022) and discuss it a bit further in the final section on future research.

## 2 Efficiency

Trials often suffer from recruitment difficulties, with estimates of 35% (between 1994 and 2002) and 56% (between 2004 and 2016) not reaching the goal set in advance (McDonald *et al.*, 2006; Walters *et al.*, 2017). These trials find themselves underpowered according to their own protocol: when they decide the stop the recruitment and obtain the final sample size for analysis, they have a high probability for their test statistic to fall outside the rejection region they set in advance, also if the effect they set out to find is there. During the COVID-19 pandemic many RCTs missed the peak of the pandemic in their region and were at risk of failing to meet recruitment targets, which motivated the proposal for 'real-time' meta-analysis (Petkova *et al.*, 2020). Unfortunately, the literature on *research waste* (Chalmers & Glasziou, 2009) and *Evidence-Based Research* (Lund *et al.*, 2016) shows that the existing evidence base is not used well to design the new trials needed for conclusion or to interpret new research. ALL-IN meta-analysis makes this very easy to do. It comes with a simple notion of the evidence already collected and what is still needed, and a notion of a new trial's ability to provide that: anticipated *e*-growth. The combination of the two has the capacity to make study design more honest, showing what a trial can add to the existing evidence base instead of just evaluating a misguided goal to single-handedly answer a research question.

## 2.1 The evidence so far and what is still needed

An ALL-IN meta-analysis can set a prospective goal for conclusion, e.g. $\alpha = 0.0025 = 0.05^2$ corresponding to the level of $\alpha$ required by authorities like the FDA that ask for two trials at the $\alpha = 0.05$ level. Following Ville's inequality (5) we need a betting score of $1/\alpha$ = €400 if we start with €1 to reach a conclusion. Because an ALL-IN meta-analysis combines trials by reinvesting or multiplying betting scores, a very simple calculation gives the betting score we still need at any given point. If an initial trial is able to reach a score of €8, any new trial can be designed to multiply that by 50. So on its own, starting with €1 instead of €8, it would need a betting score of €50 to help the meta-analysis reach €400. We could evaluate the sample size of the new trial on its ability to reach 50, which for a fixed sample size gives the conditional power of the ALL-IN meta-analysis once the new trial is added. However, if this second trial also foresees recruitment issues, it is more difficult to evaluate its planned contribution since it will probably not be the final trial in the meta-analysis. For this, we propose a continuous notion of the ability of a study: the anticipated *e*-growth. This concept was first proposed by Shafer (2021) to accompany the betting score or *e*-value with a study design property that "tells a coherent story" unlike power and *p*-values where power requires a fixed significance level while the *p*-value does not. Shafer's concept 'implied target' specifically focused on settings in which the betting strategy is not explicit because no alternative hypothesis is specified. Here we focus on an explicit alternative hypothesis and sample size anticipated in the study design, and propose the term 'anticipated *e*-growth' to evaluate what a single study is thought to add to an existing line of research. We mostly use this concept to compare the ability of various studies in a meta-analysis to each other, and judge whether new studies can conclude the meta-analysis in terms of the assumptions of their study teams. The concept of 'power' does not have that cumulative nature. In statistical literature, power can also be a confusing concept because it is used both as a property of the study and as a property of a statistical test (for statements like 'the chi-squared test has more power than the Fisher exact test). In the literature on testing by betting and *e*-values, we do not use the same term for both properties, and use 'anticipated *e*-value growth' for a property of study design, and use the term 'e-power' (Ramdas & Wang, 2024) for the property of an *e*-value test.

## 2.2 The ability of a new trial: the anticipated *e*-growth

The betting score, *e*-value or likelihood ratio summarizes the data not in just two categories—statistical significant or not statistical significant—but captures the evidence so far on its way to a certain threshold. Similarly we propose to not evaluate experimental design as all-or-nothing, but summarize its ability to build on what is already there and facilitate future research. To capture a study's expected contribution to a series of studies, we formulate the anticipated *e*-growth as the multiplicative amount with which the combined evidence is expected to grow if the study—designed with a certain $\mu_{\text{antic}}$ and sample size $n$—is added. In general, the anticipated *e*-growth $E^*$ is defined on a $Z$-statistic as follows:

$$E^* = \exp\left(\mathbf{E}_{Z^{(n)} \sim \phi_{\mu_{\text{antic}}\sqrt{n}}}\left[\log\left(\mathbf{LR}^{(n)}(Z^{(n)})\right)\right]\right). \tag{8}$$

The logarithm appears in equation (8) because the distribution of a betting score/*e*-value/likelihood ratio based on $n$ events is very non-symmetric and heavy tailed, with extremely large likelihood ratios occurring with not so small probability (see Figure 4). So the expectation of the likelihood ratio is drawn very far from its typical values by these large likelihood ratios and is not a good expression of what to expect. The logarithm makes the distribution more symmetric (asymptotically (for large $n$) and for normal likelihood ratios even normally distributed), such that the expectation is a more meaningful summary of the evidence promised by the study. By exponentiation (exp()) we bring this expectation back to the scale of the likelihood ratio, such that it can be interpreted as a betting score or *e*-value.

In the FDA COVID-19 vaccine game the expected growth rate per new event in the CureVac trial, assuming their effect of minimal interest of 60% VE, is the following:

$$\begin{aligned}
&\exp\left(\mathbf{E}_{60\%\ \mathrm{VE}}\left[\log\left(\frac{\mathcal{L}(50\%\ \mathrm{VE}\,|\,X)}{\mathcal{L}(30\%\ \mathrm{VE}\,|\,X)}\right)\right]\right)\\
&= \exp\left(\frac{40}{140}\cdot\log\left(\frac{50/150}{70/170}\right)+\frac{100}{140}\cdot\log\left(\frac{100/150}{100/170}\right)\right)\\
&= 1.029454.
\end{aligned}$$

The cumulative contribution of each new event is shown as the linear line on a logarithmic scale in Figure 5. The CureVac AG (2020, Table 8) design planned a final analysis at $n = 160$ events, so their anticipated *e*-growth was $1.029454^{160} \approx 104$. In comparison to anticipated *e*-growth of €104 at 160 events, the actual betting score €1.84 after 83 + 145 = 228 events in the press release is quite disappointing.

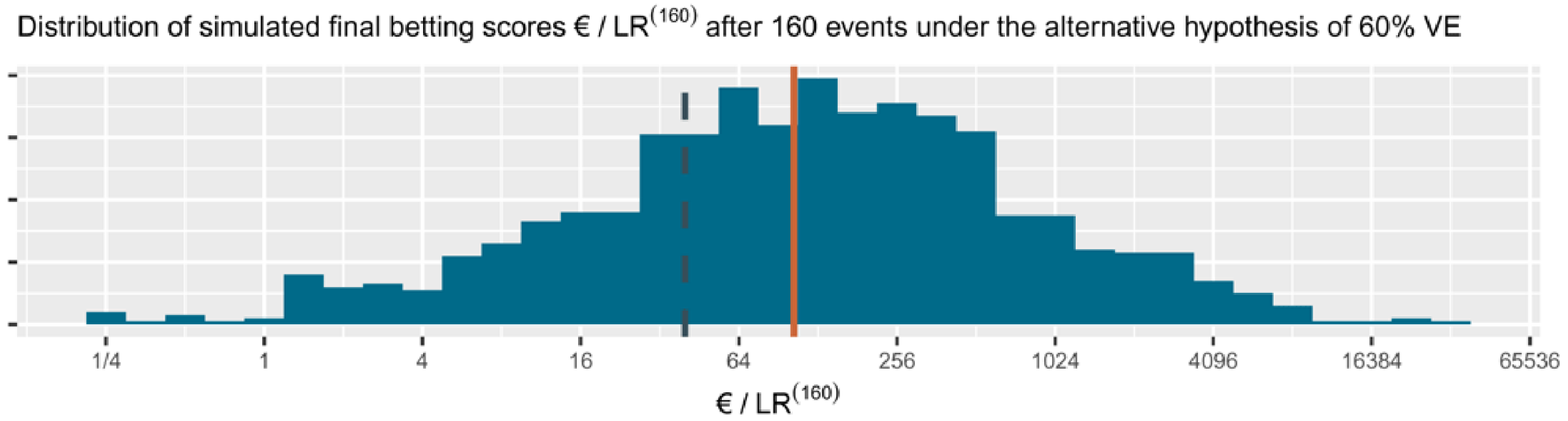

**Figure 4.** **(and Figure 5) 1000 simulated sequences of betting scores by round in the FDA COVID-19 vaccine game after 160 events assuming a probability of 0.29 (40/140) for each event to occur in the vaccine group.** This is the alternative hypothesis of 60% vaccine efficacy (VE) used to power the CureVac AG (2020) trial at a number of events of 160. The dashed line is the threshold $1/\alpha = 40$ one-sided and the solid line is the anticipated *e*-growth of €104. Note that the horizontal axis is on a logarithmic scale.

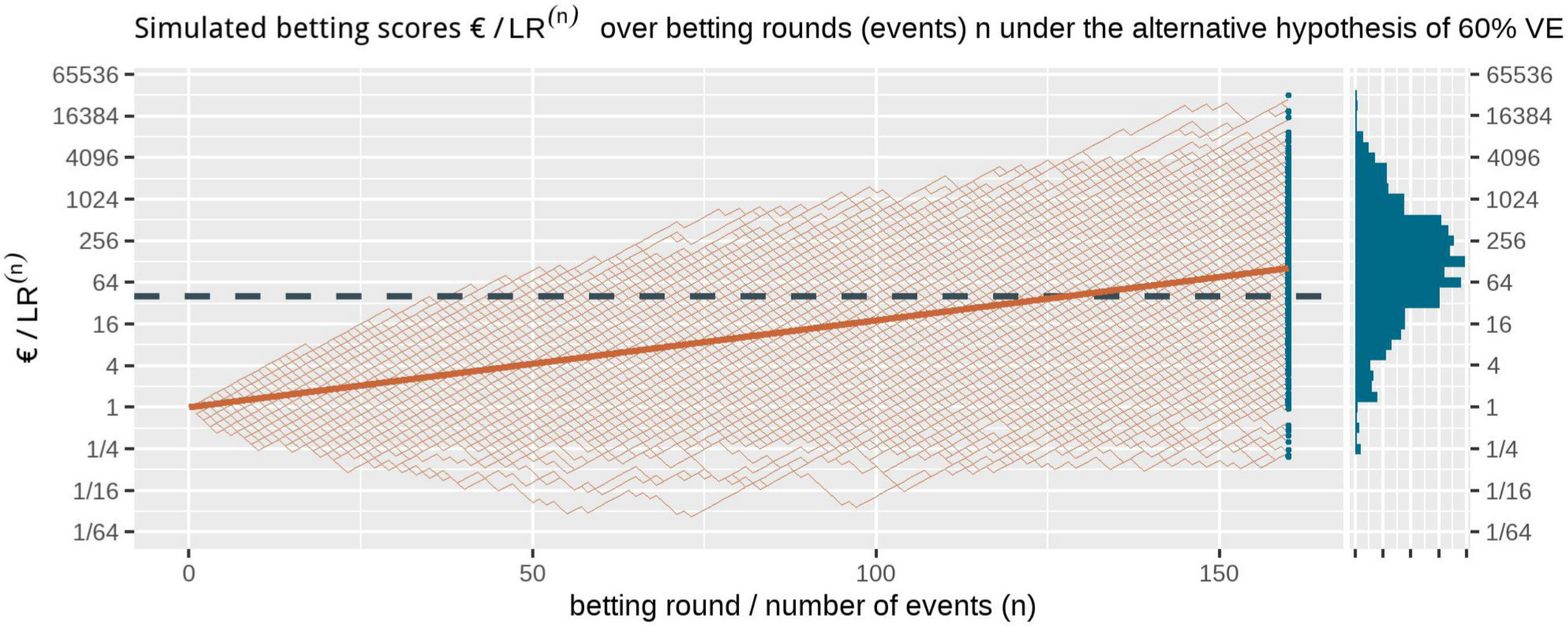

**Figure 5.** **(See above at Figure 4).** The histogram for the final betting scores at the right shows the larger scores above and the smaller ones at the bottom, which means that if we turn it, it is the mirror image of the histogram in Figure 4. The dashed line is the threshold $1/\alpha = 40$ one-sided. The increase in the solid line per additional event/betting round shows the contribution to the anticipated *e*-growth of each event, up until the anticipated *e*-growth at $n = 160$ of 104. In this figure, the design has an approximate 79% power to observe a betting score/*e*-value larger than $1/\alpha = 40$ before 160 events and 72% power at exactly 160 events (better visible in Figure 4). Note that the vertical axis is on a logartihmic scale.

### 2.3 Honest study design

An anticipated *e*-growth does require an honest proposal of the anticipated effect $\mu_{\text{antic}}$, to evaluate the merits of the study. In reality, sometimes this parameter is tweaked—e.g. setting an unrealistically large effect—to still argue for the study's advancement with only small sample size. Or the effect size of minimal clinical importance is set after data is observed (Wang *et al.*, 2018). This behavior is incentivized by the *all-or-nothing* character of Neyman-Pearson tests that also make the power analysis all-or-nothing. If your desired sample size does not meet the power hoped-for, you need to either increase it or abandon the study. This aspect of traditional analyses fully ignores the ideal of cumulative science in which one study is not expected to single-handedly answer a research question and small increments in knowledge are valuable, as long as they build towards a common goal. If they use *e*-values and the ALL-IN framework, researchers do not have to view their analysis as the final one, which helps them to design their study more honestly (Lakens, 2022).

## 3. Collaboration

The *Evidence-Based Research Network* (Lund *et al.*, 2016) aims to always inform new research by past results and to reduce research waste by separating research ideas that are necessary from those that are wasteful. This is not easy to do, however. Different communities might have different notions of necessity or even of what is ethical (a state of so-called *clinical equipoise* (Shamy *et al.*, 2020)). It might therefore be very beneficial to have all those running new clinical trials in a field collaborate together in a prospective meta-analysis. The ALL-IN approach is the most flexible way to do so and might motivate participating trials by the promise of an early conclusion if performed in 'real-time' on INterim data.

We ran two ALL-IN prospective meta-analyses during the COVID-19 pandemic with the involvement of seven trials in one and four in the other. All were designed to study whether the BCG vaccine, originally developed to protect against tuberculosis, could protect against COVID-19 (based on a theory of non-specific immune effects and innate immunity (Netea *et al.*, 2020)). The two meta-analyses study different populations (healthcare workers and the elderly) and two questions each: the effect of the BCG vaccine on COVID-19 infection (not necessarily symptomatic) and the effect on severe COVID-19 (indicated by hospitalizations). In the following illustration we will focus on the analysis of COVID-19 infections in the healthcare workers population, and the Secondary analysis that includes all trials[5].

The project was named ALL-IN-META-BCG-CORONA, and followed many of the recommendations in guidelines for prospective meta-analysis (see the publication ter Schure *et al.*, 2022 for details). It identified trials by searching clinical trial registries. It set the inclusion criteria and analysis plans in advance, before results were known. Outcome measures were harmonized among the participating trials, following consensus procedures. And all data was curated twice, by the trial statistician and the meta-analysis statistician. Apart from these best practices, there were also aspects of the project that are specific to using the ALL-IN statistical approach, which we will briefly discuss here: (1) outside control not necessary but keeping track of results in a dashboard, (2) top down or bottom up: the design of the trials and the meta-analysis design, (3) collaboration in a competitive field or a pandemic: type-I, type-II errors and publication bias, and (4) fixed-effects instead of random-effects meta-analysis.

### 3.1 Outside control not necessary but keeping track of results in a dashboard

Trial researchers might hesitate to collaborate if that means outsourcing decisions in their trial on early stopping or extending follow-up to an outside committee, such as a meta-analysis Data and Safety Monitoring Board recommended by others for 'real-time' meta-analysis (Petkova *et al.*, 2020). Instead, the ALL-IN approach allows for a 'soft' threshold that merely guides the trials; the leading source of information, but not enforcing anything. This threshold can set a stopping rule, but not in the strict sense that any rule would for a group-sequential or simulation-calibrated design. Those are much stricter because the analysis becomes invalid when the stopping rule is not enforced. In the ALL-IN approach, the threshold can inform how close to a conclusion the meta-analysis is, while also allowing individual trials to continue their follow-up based on trial-specific rationalizations. Adding new events (or even trials) after the threshold is crossed does in fact not invalidate the interpretation of the analysis.

In the ALL-IN-META-BCG-CORONA collaboration, the interim meta-analysis results were communicated through a dashboard. Access to the dashboard could be managed by adding logins and changing permissions to inspect data, and was first only given to one person per trial that was necessarily already unblinded to their own trial results: the one uploading the data for the meta-analysis. At first, each data-uploader received a dashboard account with permissions only to inspect the meta-analysis *e*-value and their own trial contribution. This access of interim meta-analysis results in the dashboard served as a motivator to keep their own trial data upload up-to-date and to check the sequence of *e*-values for errors. After an initial period where the data-uploaders could only inspect their own trial results, they granted each other permission to inspect all the individual trial contributions. When the first trials were completed and the meta-analysis was approaching its conclusion, the results were also presented to the Advisory and Steering committees. Figure 6 shows this dashboard based on a demo login with synthetic data (this demo was available for everyone involved from the start to get an impression).

A dashboard for ALL-IN meta-analysis allows to spot trends in the accumulating evidence, and allows any stakeholder to monitor the evidence to prepare for crossing a threshold in the near future, e.g. for independent data monitoring committees of ongoing trials or for those considering new trials or preparing to update medical guidelines. On a log-scale, the increase in *e*-values is linear (in expectation, see Figure 5) and the observed trends can be projected forward in time, e.g. in Figure 6 as an increase in evidence per additional calendar day.

For ALL-IN-META-BCG-CORONA, the time unit $t$ in the definition of $\mathbf{LR}^{(t)}$ from (4) was set to calendar days and the *e*-values were updated at each calendar day with an event. The dashboard plots in Figure 6 horizontal lines at 1 for trials that do not observe any events yet: they have not started betting and are still at their initial investment of €1 contributing a neutral amount to the multiplication meta- *e*-value. ALL-IN meta-analysis monitors *e*-values as events come in, also when they do so from multiple trials simultaneously. In the language of betting, even the analysis of simultaneous events is considered a sequential bet. If the bet on the events from one trial pays out €4, it multiplies our initial capital by 4, and if the events from another trial pay out €5, it does so by a factor 5. Yet if we actually consider those trials to be consecutive bets, we reinvest the €4 from the first into the second, and obtain €1 · 4 · 5 = €20, as follows from the definition of the meta-analysis *e*-value on interim data in (4).

A dashboard such as in Figure 6 encourages inspection of each individual trial's contribution to the meta-analysis. Since each trial's contribution is a simple multiplication, their components can often be conveniently spotted in the agreement of the shape of the meta-analysis and individual trial lines in a dashboard like Figure 6 (as long as not too many trials are contributing simultaneously).

[5] The primary analysis excluded the Guinea-Bissau/Mozambique (AF) trial based on a prespecified criterium on the event definition that excluded the observed infections demonstrated by serology tests in the AF trial.

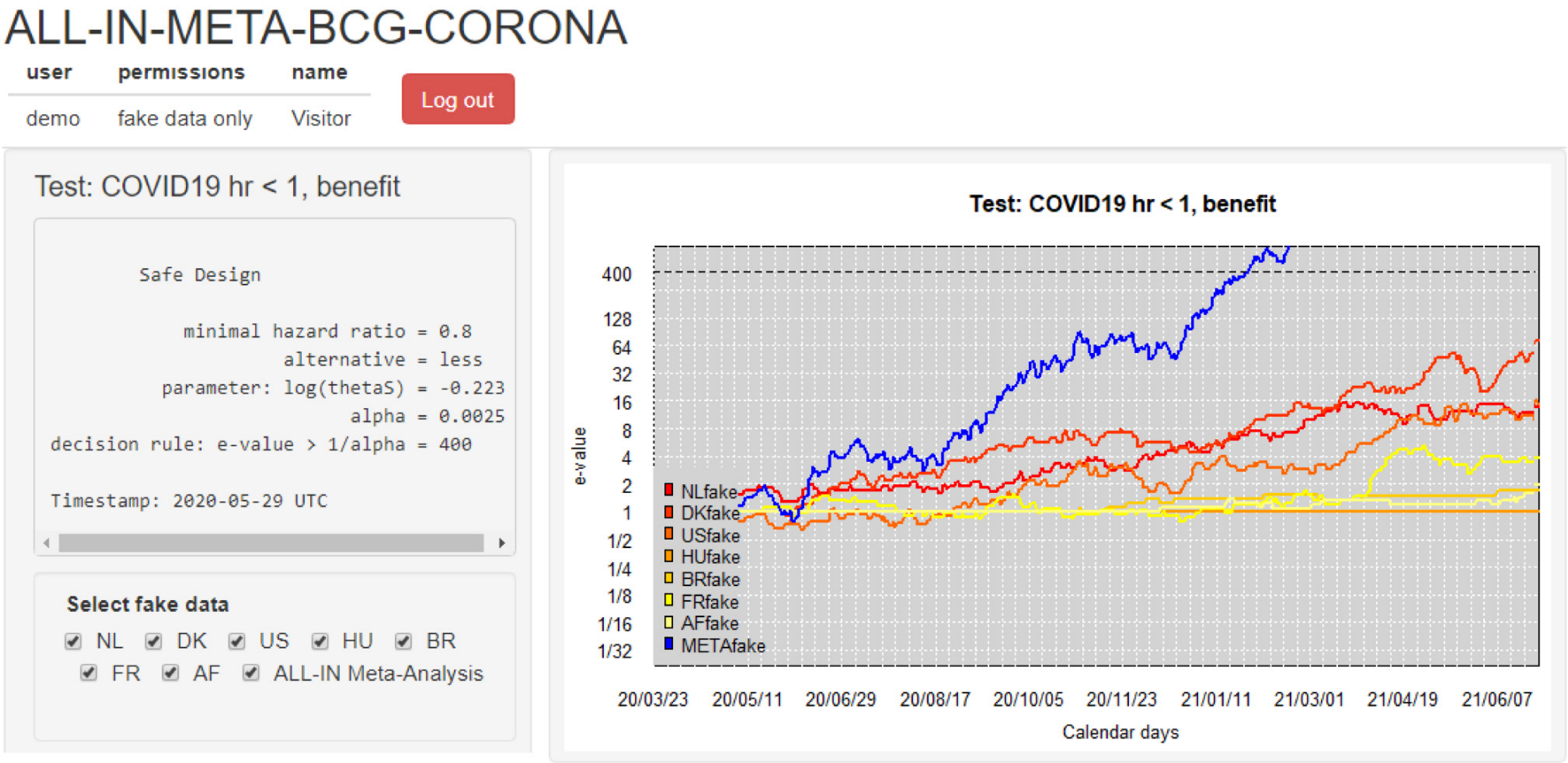

**Figure 6. Dashboard used to communicate interim results in *ALL-IN-META-BCG-CORONA* to all data uploaders with a login.** The trials were performed in the Netherlands (NL), Denmark (DK), the United States (US), Hungary (HU), Brazil (BR), France (FR) and Guinea-Bissau/Mozambique (AF)[6]. The dashboard is in demo mode and shows synthetic ("fake") data. The option to (de)select trials is for plotting purposes of individual trial *e*-values; all trials in the dashboard stay included in the meta *e*-value, following the decision from the Steering committee on trial inclusion. Note that the vertical axis is on a logarithmic scale.

[6] Datasharing for the French trial (FR) was delayed considerably by data-transfer agreements, such that this trial's results were not yet included in the publication reported in Table 1. The South African (SA) trial, on the other hand, was much faster with data sharing and therefore included, even though it was not anticipated at the time the dashboard was launched.

## 3.2 Top down or bottom up: the design of the trials and the meta-analysis design

The design of the meta-analysis was mostly top-down by a Steering Committee, but could have also included per-trial elements. In the ALL-IN-META-BCG-CORONA project it was possible, and deemed preferable, to let the Steering Committee decide on an IPD meta-analysis on interim data ('real-time') and write protocols and statistical analysis plans, still fully blinded to any results. This timestamped three important decisions on the meta-analysis design: the null hazard ratio of 1 ($VE_0 = 0\%$), the hazard ratio of minimal interest of 0.8 ($VE_1 = 20\%$) and the level of $\alpha$ set at 0.0025 so the threshold for the *e*-value was at $1/\alpha = 400$. Of these, the hazard ratio of minimal interest could have also been set differently for each trial based on individual trial designs, but not all of the trials in the collaboration were powered to detect an effect-size of minimal clinical importance specified on COVID-19 infections. So this effect-size set in a top-down manner, preregistered (Van Werkhoven *et al.*, 2021) and communicated through a webinar and newsletter and discussed in meetings, where each trial was represented by two member in the Advisory Committee. All documentation was made available on a project website (Ter Schure *et al.*, 2020a), and later added to a replication package (ter Schure, 2022a) that now accompanies the (ter Schure *et al.*, 2022) publication.

**Anticipated *e*-growth** Table 1 shows that most trials were powered for a different outcome measure[7] (indicated with '*' in the VE column) and that large effect-sizes were anticipated that were informed by observational studies. The SA trial publication, for example, reports powering 'to detect a clinically relevant target difference of 75% shown in prior BCG studies' (Upton *et al.*, 2022). Table 1 shows a simulated anticipated *e*-growth under this VE in a simulation that decreases the number of participants at risk after each observed event and calculates the exact *e*-value (see Ter Schure *et al.* (2024)). The approximation based on a stable risk set introduced for the CureVac data in Section 2.2, gives a close estimate in case the number of events is small compared to the number of participants at risk, e.g. for the US trial:

$$\exp\left(90 \cdot \mathbf{E}_{60\%\text{ VE}}\left[\log\left(\frac{\mathcal{L}(20\%\text{ VE} \mid X)}{\mathcal{L}(0\%\text{ VE} \mid X)}\right)\right]\right)$$
$$= \exp\left\{90\left(\frac{40}{140}\cdot\log\left(\frac{80/180}{100/200}\right) + \frac{100}{140}\cdot\log\left(\frac{100/180}{100/200}\right)\right)\right\} = 1.062614^{90}$$
$$= 42.3.$$

[7] unplanned absenteeism

**Table 1. Power analysis and results for the trials reported in the Secondary analysis of ALL-IN-META-BCG-CORONA, with * indicating that a trial was not powered for COVID-19 infections.** All values are obtained from openly available sources: the NL protocol publication (ten Doesschate *et al.*, 2020), the SA trial results publication (Upton *et al.*, 2022), the US clinicaltrial.gov registration (NCT04348370), the DK protocol publication (Madsen *et al.*, 2020), the HU EU Clinical Trial registration (2020-001783-28), the BR protocol publication (Junqueira-Kipnis *et al.*, 2020), the AF clinicaltrials.gov registration (NCT04641858), and the ALL-IN-META-BCG-CORONA publication and Replication Package (ter Schure *et al.*, 2022).

| | **Design (power analysis/sample size calculation)** events: COVID-19 infections | | | | **Results** events: COVID-19 infections | | | | | |
|---|---|---|---|---|---|---|---|---|---|---|
| | *anticipated e-growth given data generated under the assumption of the anticipated VE reported in the power analysis, with e-values based on $VE_1$ = 20% / $HR_1$ = 0.8 and $VE_0$ = 0% / $HR_0$ = 1.* | | | | *e-values based on $VE_1$ = 20% / $HR_1$ = 0.8 prespecified by the Steering Committee as smallest effect-size of interest and $VE_0$ = 0% / $HR_0$ = 1.* | | | | | |
| | **Anticipated VE** | **Participants** | **Events (n)** | **Anticipated *e*-growth** | **Participants** | **Events (n)** | **Excess events in BCG group: sum(O-E)** | ***e*-value exact** | **Logrank *Z*-score** | ***e*-value Gauss** |
| **NL** | * | 1500 | | | 1496 | 206 | -8.5 | 1,884 | -1.19 | 1,889 |
| **SA** | 75 | 440 | 82 | 173,5 | 1000 | 172 | 5.6 | 0,097 | 0.87 | 0,096 |
| **US** | 60 | 1800 | 90 | 43,2 | 575 | 31 | 2.4 | 0,472 | 0.88 | 0,472 |
| **DK** | * | 1500 | | | 1221 | 63 | 4.0 | 0,274 | 1.02 | 0,273 |
| **HU** | * | 950 | | | 10 | 3 | 0.1 | 0,954 | 0.17 | 0,950 |
| **BR** | 50 | 400 | 38 | 3,3 | 131 | 20 | -0.7 | 1,053 | -0.35 | 1,052 |
| **AF** | * | 1050 | | | 364 | 80 | -5.0 | 1,982 | -1.20 | 2,284 |
| **ALL-IN META Secondary analysis** | | 7640 | | >>400 | 4797 | 575 | -2.0 | 0,047 | -0.17 | 0,053 |

The anticipated *e*-growth in Table 1 show that these three trials alone were anticipated to reach the 400 threshold (173.5 · 43.2 · 3.3 >> 400) and that while the SA and US trial were expected to be more important, the BR trial would still contribute, even if small on its own. As with any study design, assuming a larger effect size (larger true VE in the power analysis) results in more convincing test statistics—and the *e*-value is a test statistic as well. But while too optimistic effect sizes result in '*nothing*' in an all-or-nothing analysis—results with wide confidence intervals and large *p*-values—a promising but inconclusive *e*-value can still contribute to the line of research when combined with new data to continue reaching a conclusion.

The US, HU, BR and AF trial in this ALL-IN-META-BCG-CORONA example did not manage to recruit their intended number of participants and observe the number of events. ALL-IN meta-analysis is ideally suited to still combine such data into a meta-analysis, since the analysis method does not rely on prespecified alpha spending and therefore requires no rule on the number of events. After all, the procedure of analyzing interim data is exactly the same as for analyzing a trial that is completed.

***E*-values** The *e*-values show that the trials collect little evidence against the global null hypothesis. Table 1 shows exact *e*-values following the methods based on Individual Participant Data proposed by Ter Schure *et al.* (2024). These *e*-values in Table 1 can also be recalculated based summary statistics, specifically the logrank Z statistic and number of events as the Gaussian *e*-value (see equation (6) in Section 1.3), as reported in Table 1. Gaussian *e*-values are likelihood ratios of Gaussians comparing the HR of minimal clinical relevance to the null based on the logrank Z statistic. For example, the *e*-value for COVID-19 infections in trial NL can be recalculated as

$$\frac{\phi\left(-1.19 \,\middle|\, \mu = \frac{1}{2}\log(0.8)\sqrt{206}\right)}{\phi(-1.19 \mid \mu = 0)} \approx 1.9.$$

Figure 7 shows that also in retrospect the *e*-values were never close to the threshold of 400. Interpreted as a likelihood ratio, *e*-values below 1 mean that the likelihood of the null (VE of 0%) is better than the likelihood of the effect-size prespecified as minimal importance (VE of 20%).

**Anytime-valid confidence intervals** The lack of support for 20% VE as the effect-size of minimal clinical relevance raises the question how well the null of 0% VE is supported compared to other small effects, which is answered by the confidence intervals shown in the forest plot of Figure 8 and running forest plot in Figure 9. These anytime-valid intervals depict all values of the hazard ratio/VE for which, if taken as the null hypothesis in an *e*-value, the *e*-value is not reaching the threshold at the 5% level, i.e. e < 1/0.05 = 20 (here we use a more lenient treshold than the 400 used in Figure 7 to reject the null early). So excluded from the interval are all values that are discredited because a successful bet was possible, resulting in a large betting score or *e*-value > 1/0.05 = 20.

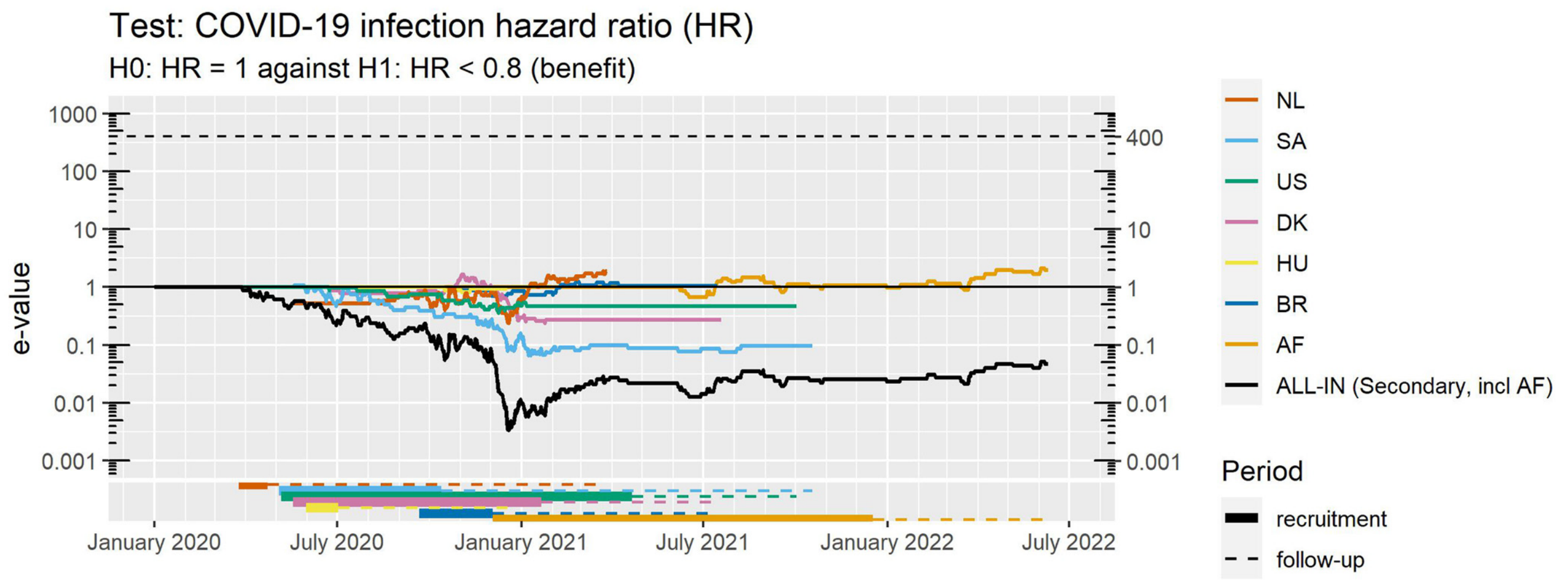

**Figure 7. Exact logrank *e*-values over time from the Secondary analysis of ALL-IN-META-BCG-CORONA with the final *e*-values also shown in Table 1.** Note that the vertical axis is on a logarithmic scale.

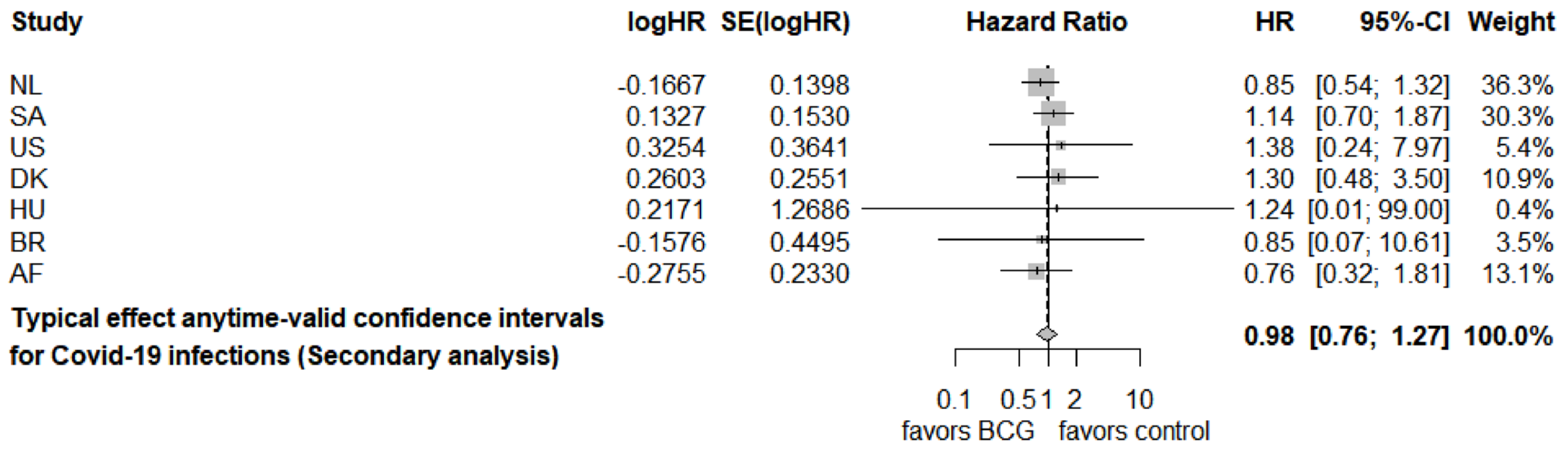

**Figure 8. Forest plot for the final Secondary analysis of ALL-IN-META-BCG-CORONA.** The figure shows intervals for the fixed-effects estimate advocated by Richard Peto (1987) as the 'typical effect'. The anytime-valid confidence interval for the HU trial is (<0.01, >99). Note that the horizontal axis is on a logarithmic scale. The hazard ratios for the trials are estimated by maximum likelihood but can also be approximated based on the excess events and total events (n) reported in Table 1 using the Peto method (Yusuf *et al.*, 1985, pp. 366-367, Statistical Appendix, see also Simmonds *et al.*, 2011) and the approximate sum of variances of the sum(O-E) assuming a stable risk set, of n·(½)·(1 – ½), e.g. for the complete ALL-IN meta-analysis: exp (sum(O–E)/sum(V) ) = exp (-2.0/ (575/4)) = 0.986.

No futility analyses were planned in this collaboration, but the 30% vaccine efficacy prescribed by the FDA for COVID-19 trials (FDA, 2020) could have been a good candidate for a futility margin and 2,5% a good significance level for the type-I error guarantee for such a strong futility analysis. Using this margin in hindsight in Figure 7, a futility conclusion could have been drawn as early as December 2020/January 2021, when none of the values in the black ALL-IN confidence interval were larger than 30% vaccine efficacy. At this point, the collaboration could have advised against starting new trials or extending recruitment in ongoing trials, and the AF trial (the last one to start recruitment around that time) might have seen that coming and been able to change course. So collaboration in ALL-IN meta-analysis can reduce research waste by providing a leading source of information for all trialists to adjust their aims and optimize the value of their trial.

### 3.3 Collaboration in a competitive field or a pandemic: type-I, type-II errors and publication bias

Prospective ALL-IN meta-analysis prevents losing type-I error control when many trials compete for answers on the same research question, e.g. in a competitive field or an uncoordinated scientific response to a pandemic. If trials are only evaluated in isolation and a response follows the first positive result of a single trial, serious multiple testing issues arise that inflate the type-I error and result in unreliable inference and, subsequently, poor decisions. This happens especially if all trials perform interim analyses on their own, and a type-I error occurs at an interim analyses before any other trial results are published to refute it.

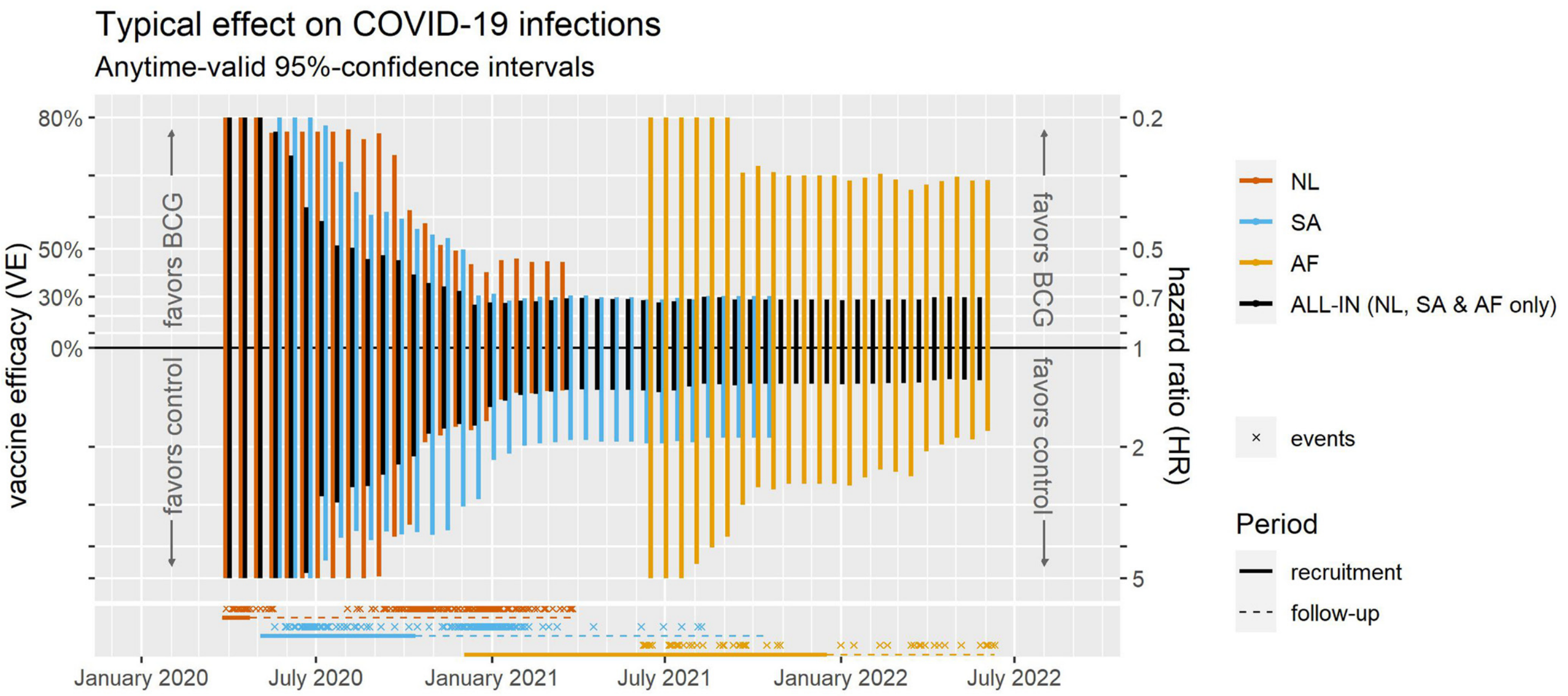

**Figure 9. Simplified sequence of forest plots for the Secondary analysis of ALL-IN-META-BCG-CORONA including only three out of seven trials for a less busy plot.** These make up 80% of the total meta-analysis weight (ter Schure *et al.*, 2022) (see all seven and their weights in the final forest plot in Figure 8). The figure shows ALL-IN meta-analysis intervals for the fixed-effects estimate advocated by Richard Peto (1987) as the 'typical effect'. Note that the vertical axis is on a logarithmic scale and that the hazard ratio scale on the right is flipped: smaller values at the top mean smaller risk in the BCG vaccinated group and therefore larger vaccine efficacy.

The example dashboard also clearly demonstrates decreased type-II errors: synthesizing the evidence in a meta-analysis at interim stages of the trials, and not after trials are completed, improves the ability to find an effect early. The COVID-19 pandemic showed that many individual studies can also be "at risk of failing to meet recruitment targets" (Petkova *et al.*, 2020). A prospective meta-analysis on interim data can "honor the participation and risk assumed by the cohorts of study participants who deserve the maximal opportunity to have their participation result in useful findings" (Petkova *et al.*, 2020).

In ALL-IN-META-BCG-CORONA, one trial had such difficulty recruiting participants, that the resulting findings ended up too noisy to merit a standalone publication (HU, see Table 1). Because the ALL-IN approach does not rely on stopping rules based on number of participants/events, this trial could still be included following the same analysis plan based on only 10 participants, as it would have with 950 participants. So recruitment difficulties do not change the position of a trial in a prospective meta-analysis and therefore do not change the standing of the research team in consensus discussions. Their knowledge and experience is optimally shared, and their data is published as part of the meta-analysis, so avoids publication bias.

### 3.4 Fixed-effects instead of random-effects meta-analysis

Sutton *et al.* (2007, p. 2491) note that "in a meta-analysis with considerable heterogeneity, the impact of a new (large) study will be (much) less in a random compared to fixed effect model". This is due the incorporation of a parameter in the model that represents the between-study variation. Also Kulinskaya & Wood (2014) find that the goal of sequentially updating a random-effect meta-analysis might involve planning a large number of small trials to estimate the between-study variance well. Even if that is considered advisable, a random-effects model result might still be very difficult to interpret (Riley *et al.*, 2011). Hence there are various reasons to prefer the fixed-effects model to monitor evidence efficiently and to ensure that the trials are sufficiently homogeneous.

Alongside the ALL-IN-META-BCG-CORONA analysis in healthcare workers we initiated a second ALL-IN meta-analysis that included trials in the elderly. Early in the process, before seeing any data, our Steering committee noticed that the two groups could be very different. Based on a theory of innate and trained immunity, they expected a different effect of the BCG vaccine on the younger immune system of healthcare workers than on the older immune system in the elderly. It could even be that the BCG vaccine effect was beneficial in the ability to fight off COVID-19 in one population but harmful in the other. In terms of statistical approach, the differences between trials can be in three categories: heterogeneous effects, conflicting effect and trial subgroup multiple testing.

**Heterogeneous effects** Our Steering committee decided that to declare success, all included trials in health-care workers should observe an effect of 20% VE or larger. If they indeed do, heterogeneity in their effect sizes (e.g. one 20%, one 50%, one 25%) does not matter for their joint ability to reject the *global null hypothesis* of no effect in all trials. So for testing the global null, trials are allowed to be heterogeneous in the space of the alternative hypothesis $H_1$ = {VE: 20% ≤ VE ≤ 100%}. For estimation, however, it is not clear what the ALL-IN confidence interval is estimating if we assume that the effects in the trials are very different. Still, as a first summary, a *typical effect size* (Peto, 1987) might be useful if we are unable to estimate a random effects model. The development of anytime-valid confidence intervals for random-effects meta-analysis is a major goal for future work. We do not, however, believe that the evidence in a line of research should be monitored based on whether this interval excludes the null hypothesis, or whether the *e*-value corresponding to the random-effects null model does: for testing, the global null is much more natural. Waiting for a random-effect model to reach a certain threshold is counter-intuitive, since it might require many small trials to estimate the between-trial variability instead of focusing on testing the treatment effect. Moreover, the goal of rejecting the null hypothesis corresponding to this model can be quite strange. When testing a zero-effect null hypothesis, it assumes that there are true effects of harm and true effects of benefit among the trials and that their mean is exactly zero.

**Conflicting effects** If one of the trials has an effect smaller than 20% or even a harmful effect, we should anticipate betting scores or *e*-values that are smaller than 1. So a meta-analysis multiplication of those *e*-values would reduce the evidence available from other trials. If we can identify groups for which we expect that the trials in each group have an effect in the same direction and of at least the minimal size, we can perform separate meta-analyses. This was the rationale behind grouping healthcare workers and the elderly each in their own ALL-IN-META-BCG-CORONA analysis.

**Trial subgroup multiple testing** When our analysis is exploratory, and we really have no idea how to group the various trials, we are faced with a multiple testing problem. Note that in this situation also no conventional meta-analysis method would be used to test a common null-hypothesis. We wonder whether any of the trials has the ability to reject the null hypothesis. In that case, we can divide our initial investment over the trials, and see if the totality of their bet achieves a high betting score. Research into this use of *e*-values has shown that indeed averaging *e*-values is the optimal way to have type-I error control in a standard multiple testing setting (Vovk & Wang, 2021) but there are many more advances in this area even improving upon *p*-value based methods for false discovery rate control and family-wise error rate control (Goeman *et al.*, 2025; Ren & Barber, 2024; Wang & Ramdas, 2022; Xu *et al.*, 2025). We return to the notion of hedging bets and averaging *e*-values in Section 4.

Problems with heterogeneity in meta-analysis are not tied to the ALL-IN approach and familiar to anyone working with meta-analysis methods. ALL-IN-META-BCG-CORONA had the advantage that many of the trials that started later had drawn inspiration from the protocol of the first trial. The same sort of alignment of inclusion criteria and outcome definitions might be achieved in other lines of research as well. Hence close collaboration can be very important and the promise of an early conclusion of the research effort might keep a research field motivated to keep the goals aligned.

## 4. Communication

We have illustrated that the language of betting can be useful in interpreting results from an ALL-IN meta-analysis. Here we argue this further by giving extensions of our method that are very easily explained in terms of betting.

### 4.1 The language of betting for two-sided tests

Our examples so far covered one-sided tests, but those can be easily extended to two-sided tests, e.g. by taking

$$\mathbf{LR}^{(n)}_{\text{two-sided}} = \frac{1}{2} \cdot \left( \mathbf{LR}^{(n)}_{\text{left}} + \mathbf{LR}^{(n)}_{\text{right}} \right),$$

with

$$\mathbf{LR}^{(n)}_{\text{left}} = \frac{\phi_{\mu_{\min(\text{left})}\sqrt{n}}(z^{(n)})}{\phi_{\mu_0}(z^{(n)})} \quad \text{and} \quad \mathbf{LR}^{(n)}_{\text{right}} = \frac{\phi_{\mu_{\min(\text{right})}\sqrt{n}}(z^{(n)})}{\phi_{\mu_0}(z^{(n)})},$$

to represent a two-sided alternative hypothesis

$$H_1 = \left\{ \phi_{\mu_1} : \ \mu_1 \le \mu_{\min(\text{left})} \ \text{ or } \ \mu_1 \ge \mu_{\min(\text{right})} \right\}.$$

Such a two-sided test is easy to interpret in the language of betting. We essentially split our initial investment (e.g. €1) between the two sides of the alternative hypothesis (e.g. by betting €0.50 on one side and €0.50 on the other). Any other weighting of the two sides is also possible and corresponds to a different division of the initial investment. The crucial thing is that each side tests the same null hypothesis $H_0$ = { $\phi_{\mu 0}$} and has expectation 1 under the null hypothesis, such that any weighted average also has expectation 1 and is an *e*-value. Note that for a meta-analysis at time $t$ with $k\langle t \rangle$ studies this becomes:

$$\mathbf{LR}^{\langle t \rangle}_{\text{two-sided}} := \frac{1}{2} \left( \prod_{i=1}^{k\langle t \rangle} \mathbf{LR}^{(n_i\langle t \rangle)}_{i,\text{left}} + \prod_{i=1}^{k\langle t \rangle} \mathbf{LR}^{(n_i\langle t \rangle)}_{i,\text{right}} \right). \tag{9}$$

Usually one side of the bet is losing and the other is winning such that we do not want to reinvest (multiply) across sides but keep them separate for all trials. In our ALL-IN-META-BCG-CORONA dashboard we also visualized these two sides of the meta-analysis test separately; in Figure 6 we show only the left-sided test (for benefit) of the two.

### 4.2 The language of betting for co-primary endpoints

Another way to hedge our bets is by considering multiple primary outcomes. In ALL-IN-META-BCG-CORONA, for example, not only were the COVID-19 events counted, but COVID-19 hospitalizations as well, as an indicator for severe disease. We started with $\alpha$ = 0.05 and put 10% on COVID-19 ($\alpha$ = 0.0025 on each of the two sides of a two-sided test) and 90% on hospitalizations ($\alpha$ = 0.0225 on each of the two sides of a two-sided test). So the thresholds to achieve with the $e$-value for COVID-19 was set at $1/\alpha = 400$ and the one for hospitalization at $1/\alpha = 44.44$. A different way to formulate this is that each had to achieve $1/\alpha = 20$, but that the sequence of $e$-values for COVID-19 started with an initial investment of €0.05 for each side of the two-sided test (and had to multiply by 400 to reach €20) and that the $e$-value for hospitalization started with an initial investment of €0.45 for each side (and had to multiply by 44.44 to reach €20).

There are two ways to consider such a bet on two co-primary outcomes: separately and combined. If we evaluate the $e$-values for each primary outcome separately and reach the threshold with either of the two, we are rejecting the null for that outcome. We are doing two separate tests. If we evaluate the $e$-values combined, we average them weighted by their $\alpha$, just as for the two sides of the two-sided test. In that case we have similar type-I error control, but reject the null hypothesis that both are a null effects in favor of the alternative hypothesis that one of them is not. Yet we cannot conclude which one is non-null with the same type-I error since our $\alpha$-level applies to the combined bet and the individual components to the averaged bet are essentially lost.

## Concluding remarks

The novelty of this paper lies in a new method for meta-analysis, and specifically all the positive practical consequences of applying the method in terms of simplicity of statistics for prospective and real-time analysis, efficiency of science (and reducing avoidable research waste), bottom-up collaboration and communication/monitoring of results. We do not claim any novelty for the underlying mathematics, though. The basic methods we describe can be viewed as relatively minor variations of the anytime-valid tests that are designed to preserve type-I error under optional stopping, as designed by H. Robbins and his students (Darling & Robbins, 1968; Robbins, 1970). Unfortunately and surprisingly, these tests have not caught on in statistics until a few years ago—right now they are thriving in work on so-called *safe tests*, *anytime-valid confidence intervals* and $e$-values e.g. Grünwald, 2024; Grünwald *et al.* (2024a); Grünwald *et al.* (2024b); Henzi & Ziegel (2022); Howard & Ramdas (2022); Howard *et al.* (2021); Koning, 2024; Pace & Salvan (2020); Ramdas *et al.* (2020); Ramdas *et al.* (2023); Shafer *et al.* (2011); Shafer (2021); Turner & Grünwald (2023); Turner *et al.* (2024); Vovk & Wang (2021); Waudby-Smith & Ramdas (2024).

This paper introduces ALL-IN meta-analysis based on Z-score methods of meta-analysis, as introduced in standard works like by Borenstein *et al.* (2009). We mainly focus on testing rather than estimation. For testing, extensions to IPD meta-analysis based on exact $e$-values (rather than Z-score approximations) follow easily since both can be combined by multiplication, as shown in our ALL-IN-META-BCG-CORONA example. Within this focus on testing with type-I error control, the heterogeneity question is less explicit, since under the global null hypothesis there is no heterogeneity, as discussed in Section 3.3. In future work we will provide more details on estimation under heterogeneity, with anytime-valid confidence intervals for the fixed-effect (singular), fixed-effects (plural) and random-effects model. Extensions to network meta-analysis seem possible, but are not yet our main direction. With regard to meta-regression, much development can be expected from work on $e$-values in settings of composite null-hypotheses, for which research is ongoing that extends to linear regression (e.g. Lindon *et al.* (2024); Pérez-Ortiz *et al.* (2024)). For IPD meta-analyses that take covariates into account, analysis of randomized controlled trials can use the so-called 'Model-X' $e$-value approach (Grünwald *et al.*, 2024b) but the non-randomized setting needs further research to develop $e$-values for Generalized Linear Mixed Models.

### Likelihood ratios, *E*-variables and *e*-values

In this paper we presented betting scores/$e$-values that are equivalent to likelihood ratios. In general though, betting scores and $e$-values are really generalizations of likelihood ratios that preserve the properties of likelihood ratios that give them a prominent role in statistics. Entire books have been written to advocate for summarizing evidence in observed data by a likelihood ratio (Edwards, 1974; Royall, 1997) and to separate the goal of measuring evidence from expressing posterior beliefs and making decisions. Likelihood ratios have the property that they can "favor a true hypothesis over a false one more and more strongly" and while a likelihood ratio can be misleading, "strong evidence cannot be misleading very often" (Royall, 1997, p. 14). This latter type-I error control is also referred to as a *universal bound* by Royall (1997) and, by recognizing Ville's inequality, can be generalized to other betting scores and $e$-values.

A betting score € is a random outcome of a bet and its random variable is an $E$-variable if it is nonnegative and for all $P \in H_0$, $\mathbf{E}_P[€] \leq 1$. For a given outcome of the bet, the value of such a random variable is the $e$-value. Ville's inequality relies on the multiplication of $E$-variables—forming a test martingale—which also has expectation smaller than 1 and thus is itself an $E$-variable. For the example $e$-values in this paper, the requirement on the expectation $\mathbf{E}_0[\mathbf{LR}] \leq 1$ holds for a simple null hypothesis, e.g. $H_0 = \{\phi_0\}$. Apart from likelihood ratios of two simple hypotheses, $e$-values can also be defined for more complicated tests—e.g. a $t$-test with a nuisance parameter for the variance—in which case the unit expectation needs to hold not for a single mean-0-normal distribution with known variance, but for all mean-0-distributions with any variance. Grünwald *et al.* (2024a) shows that it often is possible to construct $E$-variables for such composite testing problems, which is why we consider the $e$-value the right generalization of the likelihood ratio.

### Anytime-valid confidence intervals

In this paper we presented anytime-valid (AV) confidence intervals (in Figure 3, Figure 8 and Figure 9) for the hazard ratio or VE that are based on the Gaussian approximation on the maximum likelihood estimator of the hazard ratio (for details, see the *Statistical Appendix* of Ter Schure *et al.*, 2022). Research into other AV confidence intervals for the hazard ratio is still ongoing. For other estimation problems, AV confidence intervals already have been thoroughly studied, for example for medians and other quantiles (Howard & Ramdas, 2022), Gaussian means (Wang & Ramdas, 2025), odds ratios and risk differences (Turner *et al.*, 2024) and estimates of multinomial count data (Lindon & Malek, 2022). These have not, however, been extended to meta-analysis, and especially for the random-effects meta-analysis model, research into AV confidence intervals is a major goal of future work.

## Data availability

Data from ALL-IN-META-BCG-CORONA is available in the Replication package (ter Schure, 2022a).

## Software availability

The `safestats` R package (R Core Team, 2024; Turner *et al.*, 2022) provides software to do an *e*-value analysis for the *t*-test, *Z*-test, logrank test and 2*x*2-tables. Also functions are available to calculate the power and anticipated *e*-growth for these study designs. Anytime-valid confidence intervals can be calculated for the effect size in the *t*-test, the odds ratio in 2x2-tables and the hazard ratio in time-to-event data.

R code for the calculations, simulations and plots in this paper can be found on the Open Science Framework (Ter Schure, 2025, https://doi.org/10.17605/OSF.IO/D9JNY). These are based on the packages ggplot2 (Wickham, 2016), ggside (Landis, 2024), ggforce (Pedersen, 2024), dplyr (Wickham *et al.*, 2023), tibble (Müller & Wickham, 2023), stringr (Wickham, 2023), meta (Balduzzi *et al.*, 2019) and survival (Therneau, 2024).

The code that produces the dashboard in Figure 6 is not publicly available since it mainly deals with logins and permissions that only concern those involved at the time they were still (partly) blinded to the results.

## Acknowledgements

We acknowledge Henri van Werkhoven for his confidence and nerve. Amid the chaos of the initial COVID-19 pandemic months, he quickly grasped the subtleties and benefits of our ideas and committed to intensify a network of COVID-19 research to implement ALL-IN-META-BCG-CORONA. In this partnership, he structured our thinking for this paper. We further acknowledge Marc Bonten and Mihai Netea for the atmosphere of collaboration they put in place in BCG vaccine research and their public stance against “each-small-study-on-its-own” research culture. We also thank Glenn Shafer, Daniël Lakens, Muriel Pérez and Alexander Ly for feedback and extensive discussions. We are especially grateful to Alexander for being the lead developer of the *e*-value logrank test software and co-meta-statistician on ALL-IN-META-BCG-CORONA. We also appreciate Alexander’s help in coding the anticipated *e*-growth presented in Table 1 for these trials with a fixed number of events (n), within the current possibilities of the R package `safestats`. We thank our reviewers Junfeng Wang, Ewelina Rogozinska and Shubhendu Trivedi for useful comments.

## References

Akl EA, Meerpohl JJ, Elliott J, *et al.*: **Living systematic reviews: 4. Living guideline recommendations.** *J Clin Epidemiol.* 2017; **91**: 47–53.
PubMed Abstract | Publisher Full Text

Altman DG: **The scandal of poor medical research.** *BMJ.* 1994; **308**(6924): 283–284.
PubMed Abstract | Publisher Full Text | Free Full Text

Balduzzi S, Rücker G, Schwarzer G: **How to perform a meta-analysis with R: a practical tutorial.** *Evid Based Ment Health.* 2019; **22**(4): 153–160.
PubMed Abstract | Publisher Full Text | Free Full Text

Borenstein M, Hedges LV, Higgins JPT, *et al.*: **Introduction to meta-analysis.** John Wiley & Sons, Ltd, 2009.
Publisher Full Text

Branswell H: **12 lessons COVID-19 taught us about developing vaccines during a pandemic**. 2021; Accessed: 12 July 2021.
Reference Source

Breiman L: **Optimal gambling systems for favorable games.** *Fourth Berkeley Symposium.* 1961.
Reference Source

Chalmers I, Glasziou P: **Avoidable waste in the production and reporting of research evidence.** *Lancet.* 2009; **374**(9683): 86–89.
PubMed Abstract | Publisher Full Text

Chalmers TC, Lau J: **Meta-analytic stimulus for changes in clinical trials.** *Stat Methods Med Res.* 1993; **2**(2): 161–172.
PubMed Abstract | Publisher Full Text

CureVac AG: **Clinical trial protocol a phase 2b/3, randomized, observer-blinded, placebo-controlled, multicenter clinical study evaluating the efficacy and safety of investigational SARS-COV-2 mRNA vaccine CVnCoV in adults 18 years of age and older**. 2020; Accessed: 16 July 2021.
Reference Source

CureVac AG: **CureVac final data from phase 2b/3 trial of first-generation COVID-19 vaccine candidate, CVnCoV, demonstrates protection in age group of 18 to 60**. 2021; Accessed: 16 July 2021.
Reference Source

Darling DA, Robbins H: **Some nonparametric sequential tests with power one.** *Proc Natl Acad Sci U S A.* 1968; **61**(3): 804–9.
PubMed Abstract | Publisher Full Text | Free Full Text

Edwards AWF: **Likelihood: an account of the statistical concept of likelihood and its application to scientific inference.** Cambridge University Press, New York, 1974; **41**(4): 427–429.

Elliott JH, Synnot A, Turner T, *et al.*: **Living systematic review: 1. introduction—the why, what, when, and how.** *J Clin Epidemiol.* 2017; **91**: 23–30.
PubMed Abstract | Publisher Full Text

FDA: **Development and licensure of vaccines to prevent COVID-19.** 2020; Accessed: 12 July 2021.
Reference Source

Glasziou P, Chalmers I: **Research waste is still a scandal—an essay by Paul Glasziou and Iain Chalmers.** *BMJ.* 2018; **363**: k4645.
Publisher Full Text

Glasziou PP, Sanders S, Hoffmann T: **Waste in COVID-19 research.** *BMJ.* 2020; **369**: m1847.
PubMed Abstract | Publisher Full Text

Goeman J, de Heide R, Solari A: **The e-Partitioning principle of false discovery rate control.** arXiv preprint arXiv:2504.15946, 2025.
Publisher Full Text

Godolphin PJ, Rogozińska E, Fisher DJ, *et al.*: **Meta-analyses based on summary data can provide timely, thorough and reliable evidence: don't dismiss them yet.** *Nat Med.* 2022; **28**(3): 429–430.
PubMed Abstract | Publisher Full Text

Goldfeld KS, Wu D, Tarpey T, *et al.*: **Prospective individual patient data meta-analysis: evaluating convalescent plasma for COVID-19.** *Stat Med.* 2021; **40**(24): 5131–5151.
PubMed Abstract | Publisher Full Text | Free Full Text

Goudie AC, Sutton AJ, Jones DR, *et al.*: **Empirical assessment suggests that existing evidence could be used more fully in designing randomized controlled trials.** *J Clin Epidemiol.* 2010; **63**(9): 983–991.
PubMed Abstract | Publisher Full Text

Grünwald PD: **Beyond Neyman–Pearson: e-values enable hypothesis testing with a data-driven alpha.** *Proceedings of the National Academy of Sciences.* 2024; **121**(39): e2302098121.
Publisher Full Text

Grünwald P: **Peter D. Grünwald's contribution to the discussion of 'testing by betting: a strategy for statistical and scientific communication' by Glenn Shafer.** *J R Stat Soc Series A.* 2021; **184**(2): 440–441.
Publisher Full Text

Grünwald P, de Heide R, Koolen W: **Safe testing.** *J R Stat Soc Series B Stat Methodol.* 2024a; **86**(5): 1091–1128. Available on arXiv since 2019, arXiv:1906.07801.
Publisher Full Text

Grünwald P, Henzi A, Lardy T: **Anytime-valid tests of conditional independence under model-X.** *J Am Stat Assoc.* 2024b; **119**(546): 1554–1565.
Publisher Full Text

Henzi A, Ziegel JF: **Valid sequential inference on probability forecast performance.** *Biometrika.* 2022; **109**(3): 647–663.
Publisher Full Text

Howard SR, Ramdas A: **Sequential estimation of quantiles with applications to A/B testing and best-arm identification.** *Bernoulli.* 2022; **28**(3): 1704–1728.
Publisher Full Text

Howard SR, Ramdas A, McAuliffe J, *et al.*: **Time-uniform, nonparametric, nonasymptotic confidence sequences.** *Ann Stat.* 2021; **49**(2): 1055–1080.
Publisher Full Text

Hu M, Cappelleri JC, Lan KKG: **Applying the law of iterated logarithm to control type I error in cumulative meta-analysis of binary outcomes.** *Clin Trials.* 2007; **4**(4): 329–40.
PubMed Abstract | Publisher Full Text

Ioannidis JPA, Greenland S, Hlatky MA, *et al.*: **Increasing value and reducing waste in research design, conduct, and analysis.** *Lancet.* 2014; **383**(9912): 166–175.
PubMed Abstract | Publisher Full Text | Free Full Text

Jackson D, Turner R: **Power analysis for random-effects meta-analysis.** *Res Synth Methods.* 2017; **8**(3): 290–302.
PubMed Abstract | Publisher Full Text | Free Full Text

Janiaud P, Hemkens LG, Ioannidis JPA: **Challenges and lessons learned from COVID-19 trials: should we be doing clinical trials differently?** *Can J Cardiol.* 2021; **37**(9): 1353–1364.
PubMed Abstract | Publisher Full Text | Free Full Text

Junqueira-Kipnis AP, Dos Anjos LRB, Barbosa LCS, *et al.*: **BCG revaccination of health workers in Brazil to improve innate immune responses against COVID-19: a structured summary of a study protocol for a randomised controlled trial.** *Trials.* 2020; **21**(1): 881.
PubMed Abstract | Publisher Full Text | Free Full Text

Kelly JL: **A new interpretation of information rate.** *Bell System Technical Journal.* 1956; **35**(4): 917–926.
Publisher Full Text

Koning NW: **Continuous testing.** arXiv preprint arXiv: 2409.05654, 2024.

Konnikova M: **The biggest bluff: how i learned to pay attention, master myself, and win.** Penguin, 2020.
Reference Source

Kulinskaya E, Huggins R, Henry Dogo S: **Sequential biases in accumulating evidence.** *Res Synth Methods.* 2016; **7**(3): 294–305.
PubMed Abstract | Publisher Full Text | Free Full Text

Kulinskaya E, Wood J: **Trial sequential methods for meta-analysis.** *Res Synth Methods.* 2014; **5**(3): 212–220.
PubMed Abstract | Publisher Full Text

Lakens D: **Sample size justification.** *Collabra: Psychology.* 2022; **8**(1): 33267.
Publisher Full Text

Lan KG, Hu M, Cappelleri JC: **Applying the law of iterated logarithm to cumulative meta-analysis of a continuous endpoint.** *Statistica Sinica.* 2003; **13**(4): 1135–1145.
Reference Source

Landis J: **ggside: side grammar graphics**. R package version 0.3.1. 2024.
https://CRAN.R-project.org/package=ggside

Larsson M, Ramdas A, Ruf J: **The numeraire e-variable and reverse information projection.** arXiv preprint arXiv:2402.18810, 2024.
Publisher Full Text

Lau J, Schmid CH, Chalmers TC: **Cumulative meta-analysis of clinical trials builds evidence for exemplary medical care.** *J Clin Epidemiol.* 1995; **48**(1): 45–57; discussion 59-60.
PubMed Abstract | Publisher Full Text

Lee JJ, Price JC, Jackson WM, *et al.*: **COVID-19: a catalyst for transforming randomized trials.** *J Neurosurg Anesthesiol.* 2022; **34**(1): 107–112.
PubMed Abstract | Publisher Full Text

Lindon M, Ham DW, Tingley M, *et al.*: **Anytime-valid inference in linear models and regression-adjusted inference.** Harvard Business School, 2024.
Reference Source

Lindon M, Malek A: **Anytime-valid inference for multinomial count data.** *Adv Neural Inf Process Syst.* 2022; **35**: 2817–2831.
Reference Source

Lund H, Brunnhuber K, Juhl C, *et al.*: **Towards evidence based research.** *BMJ.* 2016; **355**: i5440.
PubMed Abstract | Publisher Full Text

Ly A, Turner R, Pérez-Ortiz MF, *et al.*: **R-package safestats**. Maintainer: Alexander Ly <a.ly@jasp-stats.org>, install in R by devtools::install_github("AlexanderLyNL/safestats",ref="logrank",build_vignettes=TRUE), 2021; accessed 22 September 2021.

Madsen AMR, Schaltz-Buchholzer F, Benfield T, *et al.*: **Using BCG vaccine to enhance non-specific protection of health care workers during the COVID-19 pandemic: a structured summary of a study protocol for a randomised controlled trial in Denmark.** *Trials.* 2020; **21**(1): 799.
PubMed Abstract | Publisher Full Text | Free Full Text

McDonald AM, Knight RC, Campbell MK, *et al.*: **What influences recruitment to randomised controlled trials? A review of trials funded by two UK funding agencies.** *Trials.* 2006; **7**: 9.
PubMed Abstract | Publisher Full Text | Free Full Text

Müller K, Wickham H: **tibble: simple data frames**. R package version 3.2.1. 2023.
https://CRAN.R-project.org/package=tibble

Netea MG, Giamarellos-Bourboulis EJ, Domínguez-Andrés J, *et al.*: **Trained immunity: a tool for reducing susceptibility to and the severity of SARS-COV-2 infection.** *Cell.* 2020; **181**(5): 969–977.
PubMed Abstract | Publisher Full Text | Free Full Text

Pace L, Salvan A: **Likelihood, replicability and Robbins' confidence sequences.** *Int Stat Rev.* 2020; **88**(3): 599–615.
Publisher Full Text

Pedersen T: **ggforce: accelerating 'ggplot2'**. R package version 0.4.2. 2024.
https://CRAN.R-project.org/package=ggforce

Pérez-Ortiz MF, Lardy T, de Heide R, *et al.*: **E-statistics, group invariance and anytime-valid testing.** *Ann Stat.* 2024; **52**(4): 1410–1432.
Publisher Full Text

Petkova E, Antman EM, Troxel AB: **Pooling data from individual clinical trials in the COVID-19 era.** *JAMA.* 2020; **324**(6): 543–545.
PubMed Abstract | Publisher Full Text

Peto R: **Why do we need systematic overviews of randomized trials? (Transcript of an oral presentation, modified by the editors).** *Stat Med.* 1987; **6**(3): 233–244.
PubMed Abstract | Publisher Full Text

Polack FP, Thomas SJ, Kitchin N, *et al.*: **Safety and efficacy of the BNT162b2 mRNA COVID-19 vaccine.** *N Engl J Med.* 2020; **383**(27): 2603–2615.
PubMed Abstract | Publisher Full Text | Free Full Text

Polanin JR, Williams RT: **Overcoming obstacles in obtaining individual participant data for meta-analysis.** *Res Synth Methods.* 2016; **7**(3): 333–341.
PubMed Abstract | Publisher Full Text

R Core Team: **R: a language and environment for statistical computing**. R Foundation for Statistical Computing, Vienna, Austria, 2024.
Reference Source

Ramdas A, Grünwald P, Vovk V, *et al.*: **Game-theoretic statistics and safe anytime-valid inference.** *Statist Sci.* 2023; **38**(4): 576–601.
Publisher Full Text

Ramdas A, Ruf J, Larsson M, *et al.*: **Admissible anytime-valid sequential inference must rely on nonnegative martingales.** arXiv preprint arXiv: 2009.03167. 2020.
Publisher Full Text

Ramdas A, Wang R: **Hypothesis testing with e-values.** 2024; arXiv:2410.23614 [math.ST].
Publisher Full Text

Ren Z, Barber RF: **Derandomised knockoffs: leveraging *e*-values for false discovery rate control.** *J R Stat Soc Series B Stat Methodol.* 2024; **86**(1): 122–154.
Publisher Full Text

Riley RD, Higgins JPT, Deeks JJ: **Interpretation of random effects meta-analyses.** *BMJ.* 2011; **342**: d549.
PubMed Abstract | Publisher Full Text

Robbins H: **Statistical methods related to the law of the iterated logarithm.** *Ann Math Statist.* 1970; **41**(5): 1397–1409.
Publisher Full Text

Royall R: **Statistical evidence: a likelihood paradigm**. CRC press, 1997; **71**.
Publisher Full Text

Seidler AL, Hunter KE, Cheyne S, *et al.*: **A guide to prospective meta-analysis.**

*BMJ.* 2019; **367**: l5342.
PubMed Abstract | Publisher Full Text

Shafer G: **Testing by betting: a strategy for statistical and scientific communication.** *J R Stat Soc Ser A Stat Soc.* 2021; **184**(2): 407–431.
Publisher Full Text

Shafer G, Shen A, Vereshchagin N, *et al.*: **Test martingales, Bayes factors and *p*-values.** *Statist Sci.* 2011; **26**(1): 84–101.
Publisher Full Text

Shamy M, Dewar B, Fedyk M: **Different meanings of equipoise and the four quadrants of uncertainty.** *J Clin Epidemiol.* 2020; **127**: 248–249.
PubMed Abstract | Publisher Full Text

Simmonds M, Salanti G, McKenzie J, *et al.*: **Living systematic reviews: 3. Statistical methods for updating meta-analyses.** *J Clin Epidemiol.* 2017; **91**: 38–46.
PubMed Abstract | Publisher Full Text

Simmonds MC, Tierney J, Bowden J, *et al.*: **Meta-analysis of time-to-event data: a comparison of two-stage methods.** *Res Synth Methods.* 2011; **2**(3): 139–149.
PubMed Abstract | Publisher Full Text

Smith ER, Flaherman VJ: **Why you should share your data during a pandemic.** *BMJ Glob Health.* 2021; **6**(3): e004940.
PubMed Abstract | Publisher Full Text | Free Full Text

Sutton AJ, Cooper NJ, Jones DR, *et al.*: **Evidence-based sample size calculations based upon updated meta-analysis.** *Stat Med.* 2007; **26**(12): 2479–2500.
PubMed Abstract | Publisher Full Text

Ten Doesschate T, Moorlag SJCFM, van der Vaart TW, *et al.*: **Two randomized controlled trials of Bacillus Calmette-Guérin vaccination to reduce absenteeism among health care workers and hospital admission by elderly persons during the COVID-19 pandemic: a structured summary of the study protocols for two randomised controlled trials.** *Trials.* 2020; **21**(1): 481.
PubMed Abstract | Publisher Full Text | Free Full Text

Ter Schure J: **Replication Package ALL-IN-META-BCG-CORONA.** *Research Equals*. 2022a.
Publisher Full Text

Ter Schure J: **Code for paper ALL-IN meta-analysis: breathing life into living systematic reviews and prospective meta-analysis**. 2025.
http://www.doi.org/10.17605/OSF.IO/D9JNY

Ter Schure J: **ALL-IN meta-analysis**. PhD thesis, Leiden University, 2022b.
https://ir.cwi.nl/pub/31587

Ter Schure J, Grünwald P: **Accumulation Bias in meta-analysis: the need to consider *time* in error control [version 1; peer review: 2 approved].** *F1000Res.* 2019; **8**: 962.
PubMed Abstract | Publisher Full Text | Free Full Text

Ter Schure JA, Ly A, Belin L, *et al.*: **Bacillus Calmette-Guérin vaccine to reduce COVID-19 infections and hospitalisations in healthcare workers – a living systematic review and prospective ALL-IN meta-analysis of individual participant data from randomised controlled trials.** *medRxiv.* 2022.
Publisher Full Text

Ter Schure J, Ly A, Pérez-Ortiz MF, *et al.*: **Safestats and ALL-IN meta-analysis project page**. 2020a.
Reference Source

Ter Schure J, Pérez-Ortiz MF, Ly A, *et al.*: **The anytime-valid logrank test: error control under continuous monitoring with unlimited horizon.** *N Engl J Stat Data Sci.* 2024; **2**(2): 190–214.
Publisher Full Text

Therneau T: **A package for survival analysis in R**. R package version 3.7-0. 2024.
https://CRAN.R-project.org/package=survival

Thomas J, Askie LM, Berlin JA, *et al.*: **Chapter 22: Prospective approaches to accumulating evidence.** In: J.P.T. Higgins, J. Thomas, J. Chandler, M. Cumpston, T. Li, M.J. Page, V.A. Welch (editors) *Cochrane Handbook for Systematic Reviews of Interventions version 6.4.* Cochrane, Updated August 2023.
Reference Source

Tierney JF, Fisher DJ, Vale CL, *et al.*: **A framework for prospective, adaptive meta-analysis (FAME) of aggregate data from randomised trials.** *PLoS Med.* 2021; **18**(5): e1003629.
PubMed Abstract | Publisher Full Text | Free Full Text

Turner RJ, Grünwald PD: **Exact anytime-valid confidence intervals for contingency tables and beyond.** *Stat Probab Lett.* 2023; **198**: 109835.
Publisher Full Text

Turner RJ, Ly A, Grünwald PD: **Generic E-variables for exact sequential *k*-sample tests that allow for optional stopping.** *J Stat Plan Inference.* 2024; **230**: 106116.
Publisher Full Text

Turner R, Ly A, Pérez-Ortiz MF, *et al.*: **safestats: Safe Anytime-Valid Inference.** R package version 0.8.6, Maintainer: Alexander Ly <a.ly@jasp-stats.org>, 2022.
Reference Source

Upton CM, van Wijk RC, Mockeliunas L, *et al.*: **Safety and efficacy of BCG re-vaccination in relation to COVID-19 morbidity in healthcare workers: a double-blind, randomised, controlled, phase 3 trial.** *EClinicalMedicine.* 2022; **48**: 101414.
PubMed Abstract | Publisher Full Text | Free Full Text

van Haren FMP, Richardson A, Yoon HJ, *et al.*: **INHALEd nebulised unfractionated HEParin for the treatment of hospitalised patients with COVID-19 (INHALE-HEP): protocol and statistical analysis plan for an investigator-initiated international metatrial of randomised studies.** *Br J Clin Pharmacol.* 2021; **87**(8): 3075–3091.
PubMed Abstract | Publisher Full Text

van Werkhoven CH, Ter Schure J, Bon-ten M, *et al.*: **Anytime Live and Leading Interim meta-analysis of the impact of Bacillus Calmette-Guérin vaccination in health care workers and elderly during the SARS-COV-2 pandemic (ALL-IN-META-BCG-CORONA)**. 2021.
Reference Source

Ville J: **Etude critique de la notion de collectif.** *Bull Amer Math Soc.* 1939; **45**(11): 824.
Reference Source

Vovk V, Wang R: **E-values: calibration, combination, and applications.** *Ann Stat.* 2021; **49**(3): 1736–1754.
Publisher Full Text

Walters SJ, Bonacho Dos Anjos Henriques-Cadby I, Bortolami O, *et al.*: **Recruitment and retention of participants in randomised controlled trials: a review of trials funded and published by the United Kingdom Health Technology Assessment Programme.** *BMJ Open.* 2017; **7**(3): e015276.
PubMed Abstract | Publisher Full Text | Free Full Text

Wang H, Ramdas A: **Anytime-valid *t*-tests and confidence sequences for Gaussian means with unknown variance.** *Seq Anal.* 2025; **44**(1): 56–110.
Publisher Full Text

Wang R, Ramdas A: **False discovery rate control with e-values.** *J R Stat Soc Series B Stat Methodol.* 2022; **84**(3): 822–852.
Publisher Full Text

Wang MQ, Yan AF, Katz RV: **Researcher requests for inappropriate analysis and reporting: a U.S. survey of consulting biostatisticians.** *Ann Intern Med.* 2018; **169**(8): 554–558.
PubMed Abstract | Publisher Full Text

Waudby-Smith I, Ramdas A: **Estimating means of bounded random variables by betting.** *J R Stat Soc Series B Stat Methodol.* 2024; **86**(1): 1–27.
Publisher Full Text

Wickham H: **ggplot2: elegant graphics for data analysis**. Springer-Verlag New York, 2016.
Reference Source

Wickham H: **stringr: simple, consistent wrappers for common string operations**. R package version 1.5.1. 2023.
https://CRAN.R-project.org/package=stringr

Wickham H, François R, Henry L, *et al.*: **dplyr: a grammar of data manipulation**. R package version 1.1.4. 2023.
https://CRAN.R-project.org/package=dplyr

Xu Z, Fischer L, Ramdas A: **Bringing closure to FDR control: beating the e-Benjamini-Hochberg procedure.** arXiv preprint arXiv:2504.11759, 2025.
Publisher Full Text

Young C, Horton R: **Putting clinical trials into context.** *Lancet.* 2005; **366**(9480): 107–108.
PubMed Abstract | Publisher Full Text

Yusuf S, Peto R, Lewis J, *et al.*: **Beta blockade during and after myocardial infarction: an overview of the randomized trials.** *Prog Cardiovasc Dis.* 1985; **27**(5): 335–371.
PubMed Abstract | Publisher Full Text

# Open Peer Review

## Current Peer Review Status: ? ? ✓

**Version 1**

Reviewer Report 21 October 2022

✓

**Shubhendu Trivedi**
Massachusetts Institute of Technology (MIT), Cambridge, MA, USA

The paper presents a new method for meta-analysis, motivated partly by (and for "breathing life" into) living systematic reviews that are used in the clinical domain which provide recommendations to prevent research waste. The authors christen their method to be ALL-IN meta-analysis, which is Anytime Live and Learning Interim Meta-Analysis. "Anytime" meaning analysis can be updated at any time and can control for type I error irrespective of any other decision making along the way. "Live" allows for a bottomup collaboration of different trials; a trial can be initiated in any way, and we can include data from the meta-analysis itself. "Interim" permits for a combination of data from trials that are still ongoing.

The paper begins with a topical motivation from the covid19 pandemic, while emphasizing that the methodology could help better evidence combination, collaboration, and communication during later pandemics, or even smaller clinical trials. Using a single trial and specifications issues by the FDA for the covid19 vaccine trails (regarding vaccine efficiency and evidence against a null hypothesis say 30% VE) , the general betting game that is the language central to much of the contribution of the paper is introduced. It is shown that the same can also be written in terms of likelihood ratios and examples for scores are calculated for a Pfizer trial and a CuraVac trial.

The betting based methodology allows the statistical analysis to not simply be all or nothing (like p testing). In the all or nothing setting, we can not continue from one trial to another without violating type I error rates, while in the betting (ALL-IN) setting one can simply update patients later on. This also permits for better efficiency (we can understand the number of participants needed to answer a research question) and collaboration (since we can combine analysis as data becomes available). The language of betting also can be interpreted in various equivalent ways (likelihood ratios, conservative p values, e values) that also allow for easy and crisp communication about the analysis.

The intuitions of the language of betting are made more precise using standard tools in the literature (Markov's inequality, and Ville's inequality). Further, the betting score underlying the test is an e-value which further permits statistical analysis (using the tools cited). logrank Z statistics are used as a running example for meta-analysis on the summary statistics -- we can simply collect the Z statistics Z_i from each trial, which can easily be combined (as shown in equation 3). Notably if we add in interim data, then from the perspective of Ville's inequality, they are indistinguishable for testing. The methodology also allows for combining data from trials without requiring a common design -- this can easily done by deciding upon a min mu parameter for each trial, using which one can still get a valid combination of different trials with valid type I guarantees. Further, method not only captures whether an effect is statistically significant or not, it also captures evidence up till now. The language of the "implied target" of Shafer is used to make this precise, which in turn can also be used to quantify how much will the evidence change if a new study with some mu and N is added.

The paper also reports testing the methodology during the covid19 pandemic in two meta-analysis. One involving 7 trials, and the other involving 4 trials - considering different populations (healthcare workers, and the elderly). The trials involved testing if BCG could help with COVID19 immunity. The results are discussed while discussing issues (and recommendations) for meta-analysis design, systematic search for trials, systematic reviews for trial inclusion, data upload, and disseminating results.

In general, I found the paper very well written. The methodology is described very clearly, along with a glimpse of the underlying statistical tools available. The advantages and recommendations that the methodology has/implies are also discussed in detail. The underlying mathematics for testing is standard, but as far as I understand this is the first application for it in the setting considered in the paper. I would recommend the paper for acceptance.

**Is the rationale for developing the new method (or application) clearly explained?**

Yes

**Is the description of the method technically sound?**

Yes

**Are sufficient details provided to allow replication of the method development and its use by others?**

Yes

**If any results are presented, are all the source data underlying the results available to ensure full reproducibility?**

No source data required

**Are the conclusions about the method and its performance adequately supported by the findings presented in the article?**

Yes

***Competing Interests:*** No competing interests were disclosed.

***Reviewer Expertise:*** Machine Learning, Statistics, Conformal Prediction, Sequential Testing, Drug Discovery.

**I confirm that I have read this submission and believe that I have an appropriate level of expertise to confirm that it is of an acceptable scientific standard.**

Author Response 03 Mar 2025

**Judith ter Schure**

We apologize for this very late reply. However late, we still hope to contribute to a complete and accurate scientific literature with a Version 2 of our paper that incorporates all the very useful comments made by the reviewers. Version 2 also further illustrates the unique features of the method by extending the example ALL-IN prospective meta-analysis in the paper with a reflection on the collaboration, summary of results and references to a complete replication package.

We thank Shubhendu for his very kind and thoughtful reflection on our paper.

We actually like '*Anytime Live and Learning Interim Meta-Analysis*' as a separate interpretation of the ALL-IN acronym, even though this might have been a typo.

In this paper we introduced the ALL-IN method by examples in which a smallest effect-size of interest could be set: the FDA COVID-19 vaccine game and the BCG collaboration. While this setting in ubiquitous in theory –with every power analysis setting a maximum sample size controlling type-II errors by pretending to be based on such an effect-size of minimal interest, in practice many clinical researchers have difficulty setting one. Moreover, in noninferiority settings it is even impossible to set one. One of us has discussed these two issues briefly in a discussion piece (Ter Schure, 2024).

While the ALL-IN methods presented in our paper are optimal in both simplicity and efficiency in settings with a smallest effect-size of interest, these other settings need further research, as well as their own introduction paper. As these settings require to *learn* the betting strategy, or alternative hypothesis parameter, from the data, we might as well call it Anytime Live and Learning meta-analysis!

ter Schure, J. (2024). Judith ter Schure's contribution to the Discussion of 'Safe testing' by Grünwald, de Heide, and Koolen. *Journal of the Royal Statistical Society Series B: Statistical Methodology*, *86*(5), 1157-1159.

***Competing Interests:*** No competing interests were disclosed.

Reviewer Report 16 September 2022

?

**Ewelina Rogozinska**

MRC Clinical Trials Unit, Institute of Clinical Trials and Methodology, University College London, London, UK

The work by ter Schure and Grünwald outlines an alternative approach to evidence synthesis, aiming to include emerging evidence in real time without increasing type-1 error. The authors suggest that their ALL-IN method can "breathe life into living systematic reviews, through better and simpler statistics, efficiency, collaboration and communication". As much as I agree with the three final points and fully support this notion, I have some doubts regarding the professed 'simplicity' of the proposed statistical approach, implementation, and generalizability of the method.

Firstly, the approach has been developed and tested only in one very unusual setting (Covid-19 pandemic), in which the accumulation of evidence over a short space of time was extreme and sharing of data and collaboration was greater than witnessed in previous years. According to Heinze *et al.* (manuscript in submission) and their proposed framework of four phases of methodological research in biostatistics, the ALL-IN method would be classified as a method in a second phase of its development. Consequently, it requires further evaluation in a range of settings and refinement before it could be considered as a viable alternative to other available methods. This should be discussed in their paper, with declarations more balanced to reflect the single setting in which their method was applied.

Secondly, contrary to the authors' claim that the introduction of terminology from game theory makes it easier to communicate the uncertainties, I am finding the sections using betting language difficult to follow. The evidence synthesis community still to some extent grapples with more standard methods of advanced evidence synthesis (Wang *et al.* BMJ 2021; 373: n736).[1] Thus, the introduction of new concepts (or their reintroduction) should be carefully thought through. Overall, I feel the manuscript would benefit from limiting the use of references to betting and investments to an essential minimum.

Finally, the authors present a real-life example of their method using BCG vaccines trials for Covid-19. The presented example resembles an approach more akin to prospective individual participant data (IPD) meta-analysis than a living systematic review or also the referenced FAME approach - both relying on aggregate rather than individual participant data. The challenges associated with accessing IPD, the non-standard approach to data analysis and lack of clear proof of its benefits push this method toward "interesting" developments rather than "a new way forward". The authors should explain more clearly how their method could help aggregate-level evidence synthesis or refocus the scope to IPD based projects. Furthermore, it would be interesting to learn how the ALL-IN compares to commonly used methods in terms of efficiency and reliability of obtained results.

Concluding, the presented method is an interesting approach to evidence synthesis; however, at the current stage of its development, it requires further evaluation of its utility for the evidence synthesis to be able to bet on it.

**References**

1. Wang H, Chen Y, Lin Y, Abesig J, et al.: The methodological quality of individual participant data meta-analysis on intervention effects: systematic review. *BMJ*. 2021. Publisher Full Text

**Is the rationale for developing the new method (or application) clearly explained?**
Yes

**Is the description of the method technically sound?**
Partly

**Are sufficient details provided to allow replication of the method development and its use by others?**
Yes

**If any results are presented, are all the source data underlying the results available to ensure full reproducibility?**
No source data required

**Are the conclusions about the method and its performance adequately supported by the findings presented in the article?**
Partly

***Competing Interests:*** No competing interests were disclosed.

***Reviewer Expertise:*** Conduct of systematic reviews and methodological reviews, IPD meta-analysis, bias in evidence synthesis, use of evidence synthesis in clinical practice guidelines,

**I confirm that I have read this submission and believe that I have an appropriate level of expertise to confirm that it is of an acceptable scientific standard, however I have significant reservations, as outlined above.**

Author Response 03 Mar 2025

**Judith ter Schure**

We apologize for this very late reply. However late, we still hope to contribute to a complete and accurate scientific literature with a Version 2 of our paper that incorporates all the very useful comments made by the reviewers. Version 2 also further illustrates the unique features of the method by extending the example ALL-IN prospective meta-analysis in the paper with a reflection on the collaboration, summary of results and references to a complete replication package.

We thank Ewelina Rogozinska for her review and for supporting our take on efficiency, collaboration and communication. We also thank her for expressing her concerns about our hopeful, but perhaps naïve, statements of simplicity in the setting of sequential meta-analysis. We indeed do not have any empirical evidence that what we propose is simple to understand. What we do know is that other available approaches to control type-I error rates in living systematic reviews have proven to be very difficult (Simmonds et al, 2017), hardly ever used and even, for general use, were recommended against by a scientific committee advising Cochrane on the matter in 2018 (reference below).

Our main argument for simplicity is nicely pointed out by reviewer Shubhendu Trivedi, writing '*we can simply collect the Z statistics Z_i from each trial, which can easily be combined (as shown in equation 3). Notably, if we add in interim data, then from the perspective of Ville's inequality, they are indistinguishable for testing.*' So ALL-IN meta-analysis needs the same summary statistics as default Z-score approaches in conventional meta-analysis, and in terms of implementation that is indeed not simpler, but also not harder than conventional meta-analysis. The major point is that the procedure is the same whether the meta-analysis is the first analysis, or whether it is the $10^{th}$ update in a living systematic review, or whether it is the $100^{th}$ update in a prospective meta-analysis of interim results of ongoing trials. The latter nicely summarized by Rogozinska as '*including emerging evidence in real time without increasing type-1 error*'. There is no other statistical approach in use in meta-analysis for which retaining type-1 error control is that simple*. Other methods need a maximum sample size, an alpha-spending function, a stopping rule to enforce on the trials, and we are convinced that these are considerations that are a lot less simple than what we propose.

We have added a section to the Version 2 of the paper making explicit that this lack of sample size restrictions is a unique feature of ALL-IN meta-analysis that does not hold for any other statistical approach in use. The introduction section on Statistics now states (references in the paper):

*Bottom-up living meta-analysis*
*The lack of restrictions on sample size is a unique feature of ALL-IN meta-analysis that sets it apart from other statistical approaches to living systematic reviews , 'adaptive' (Tierney et al., 2021) or 'real-time' (Petkova et al, 2020) prospective meta-analysis and 'metatrials' (Van Haren et al, 2020). These other approaches require a maximum sample size or a maximum number of studies to guarantee type-I error control and interval coverage for all updates of the meta-analysis. The crucial difference with ALL-IN is that for those methods to be valid, participating in a prospective meta-analysis requires outside control over a trial's data collection. Outside control is needed in the case of the 'Framework for prospective adaptive meta-analysis' (FAME) where a single sample size is set comprised of data from all ongoing trials, possibly at an interim of some of them, and single-analysis statistics is used. The consequence is that the intervals and p-values reported are only valid if never updated, so the analysis assumes that data collection stops after that point. No new analyses are valid after that maximum sample size, since all alpha is spent and possibly accumulation bias is introduced (Ter Schure & Grünwald, 2019). Similarly, outside control is needed in the operationalization of group-sequential or alpha-spending methods that need the information size relative to the final analysis (Simmonds et al, 2017), and therefore also need to enforce such maximum sample size for the reported results to be valid. Finally, even in in simulation-calibrated Bayesian approaches restrictions on interims (often) and maximum sample size (always) are imposed for the simulations to terminate. So the frequentist operating characteristics (type-I error, coverage of intervals) of the analysis are not known if trials wish to continue data collection after that maximum sample size and analyze again, or in other ways deviate from the simulated scenarios. Hence all available methods in living systematic reviews and prospective meta-analysis are by design quite top-down, "using preestablished stopping rules for safety, efficacy, futility, and harm" (Petkova et al, 2020).*

With regard to simplicity of implementation, we have decided to extend Section 3 on Collaboration with a complete worked-out and reproducible example of an ALL-IN meta-analysis. This shows, among other things, that the simplicity of the statistics strengthens the collaborations when participating trials are faced with difficulties that would otherwise invalidate the analysis. Examples of these include, recruitment difficulties, delays in starting the trial, and meta-analysis results that might make a trial change course and directly influence the future sample size in an unforeseen way.

We will address the first, second and final point in the order given. We hope that by proposing some adjustments to the paper in reply to each of his comments, we can convince Ewelina Rogozinska that the description of the method is technically sound (question 2, now answered by 'Partly') and that the conclusions about the method and its performance are adequately supported by the findings presented in the article (question 5, now answered by 'Partly'). Hence we hope our V2 will be accepted without any further reservations.

Firstly

We wrote our paper from the hopeful perspective that many good things in science that came from the COVID-19 pandemic are here to stay, like an increase in prospective meta-analysis, data sharing and standardization of e.g. outcome measures and diagnostics. We also wrote our paper for an audience that agrees that they should. But there is nothing in the ALL-IN method that would prevent its use in a retrospective setting of meta-analysis based on summary statistics from published papers. The presentation based on $Z$-scores, and the worked-out example that we added in Section 3 show that, even though we favor IPD meta-analysis or collaboratively sharing summary statistics (the exact $e$-value itself for example), an approximate analysis often is already very good and flexible. We added the following text to the abstract:

"*Because the intention of the analysis does not change the validity of the results, the results of the analysis can change the intentions ('optional stopping' and 'optional continuation' based on the results so far). On the one hand: any analysis can be turned into a living one, or even become 'real-time' by updating with new trial data or even including interim data from trials that are still ongoing – without any changes in the cut-offs for testing or the method for interval estimation. On the other hand: no stopping rule needs to be enforced for the analysis to remain valid, so participating in a prospective meta-analysis does not require outside control over data collection.*"

This characteristic of ALL-IN meta-analysis might even help further popularize these good things in science that came from the pandemic.

Secondly

We wrote our paper for an audience that considers sequential statistics to be complicated. We believe that is still our main audience because most clinical trials are not analyzed with interim analyses and most meta-analyses updates don't take into account that repeated analyses invalidate both the *p*-value and the confidence intervals reported. While spending the better part of the past ten years on sequential analysis, we ourselves consider most of sequential statistics to be complicated. This is why we think that our audience would be confused if we just provided a 'cook-book' of ALL-IN methods and simply state that type-I error control and coverage of intervals is guaranteed, while using the same method over and over again. We believe that a strong intuition is needed for an audience to accept that this can be done without any corrections to the method. The game theory terminology ís that strong intuition.

While absent in most of applied statistics today, this intuition has always been around in the history of probability theory (e.g. the work of Christiaan Huygens on games in the 17$^{th}$ century), and sequential statistics (e.g. George Alfred Barnard's discussion of Wald's probability ratio test in 1947). In the currently thriving literature on *e*-values and anytime-valid confidence intervals in mathematical statistics, the intuition of games is present almost everywhere and serves both as intuition to present existing results and as a very fruitful analogy that drives new results, see for example the overview article by Ramdas, Grünwald, Vovk and Shafer (2023). However, we agree that a good discussion is possible on whether references to betting and investments benefit intuition about statistics in general. We think they can, but as you can see from the discussion following Glenn Shafer's JRSSA paper about is, there is plenty of disagreement. Yet for intuition about the sequential nature of ALL-IN meta-analysis, we do feel that the references to betting are essential for intuition, and the method should not be presented to readers familiar with the difficulties of sequential analysis without providing this intuition.

Finally

Our extension of the BCG example now shows more clearly that the ALL-IN method can be used on IPD (based on exact *e*-values), or can be used on summary statistics (in this example: the logrank Z-score and number of events). In fact, liaising with trials to share summary statistics can also be very fruitful just like in the FAME approach (e.g. sharing the exact e-value per trial). All the analyses of the BCG example can be reproduced based on openly available summary statistics. We hope Ewelina Rogozinska agrees that this illustrates how aggregate-level evidence synthesis is just as promising with ALL-IN methods as IPD based projects.

We agree that future work should compare ALL-IN to other methods. For this introduction paper, we now focus on the settings where such a comparison is of less concern, as we write in the abstract: if the analysis changes the intentions. Take for example the situation in which a research team decides to update a meta-analysis and commit to a living systematic review, and after a first update decides to liaise with future trials in a prospective meta-analysis for future updates on interim results. In that situation there simply is no other statistical technique that guarantees type-I error control and confidence interval coverage, because 'optional continuation' decisions are made that are not prespecified and that no stopping rule-based method can handle. Hence ALL-IN breathes life into living systematic reviews in a way that no other method can.

Cochrane Scientific Committee Schmid, C., Senn, S., Sterne, J., Kulinskaya, E., Posch, M., Roes, K., and McKenzie, J. (2018). Should Cochrane apply error-adjustment methods when conducting repeated meta-analyses?

*It can be that simple in a fully subjective Bayesian analysis, but we have not seen any of those in use. Bayesian analysis in practice uses default priors (and possibly also informative priors, but never exclusively) and checks for frequentist operating characteristics by simulations.

Aaditya Ramdas, Peter Grünwald, Vladimir Vovk, Glenn Shafer "Game-Theoretic Statistics and Safe Anytime-Valid Inference," Statistical Science, Statist. Sci. 38(4), 576-601, (November 2023)

Glenn Shafer, Testing by Betting: A Strategy for Statistical and Scientific Communication, *Journal of the Royal Statistical Society Series A: Statistics in Society*, Volume 184, Issue 2, April 2021, Pages 407–431, https://doi.org/10.1111/rssa.12647

***Competing Interests:*** No competing interests were disclosed.

Reviewer Report 19 August 2022

?

**Junfeng Wang**
Division of Pharmacoepidemiology and Clinical Pharmacology, Utrecht Institute for Pharmaceutical Sciences, Utrecht University, Utrecht, The Netherlands

In this manuscript, ter Schure and Grünwald proposed a new form of meta-analysis, namely ALL-IN meta-analysis, which can facilitate living systematic reviews.

The authors provided a different understanding of evidence synthesis, and a novel way of performing meta-analyses.

The new method is illustrated in the language of betting, which is complement with the acronym of ALL-IN.

Major comments:

1. The calculation of probability of next event in group X (e.g. in page 3, the fraction of 0.41 of COVID-19 events to occur in the vaccine group, 0.41=70/(100+70)), is only valid when the number of participants is infinite in each group.

When the number of participants still at risk in one group decreases, the probability of next event occurring in this group will also decrease. So the constant probability assumption may not be valid in real practice.

2. In page 14 and 15, the authors mentioned heterogeneity several times, however, the authors avoid giving a direct answer or solution to heterogeneous results from primary studies. This should be addressed or at least extensively discussed (as a limitation).

3. The COVID-19 vaccine trials are used as an example. However, all these trials are conducted recently and in a short time period (compared to other treatment). How is the generalizability of this new methods in other treatments?

4. The method is easily extended to IPD meta-analysis. How well this method can be extended to network meta-analysis of multiple treatments?

5. How can covariates, both on study level and individual level, be adjusted in this new framework?

Minor comments:

1. In Page 7, section 1.1, the equation "70/170*170/70 + 100/170*170/100 = 1", seems not correct, the right side should be 2. But I assume the authors forget adding something in the left side of the equation. Please check.

2. "Betting on vaccine" in this paper actually means betting on the next event will occur in the vaccine group (which means vaccine is not effective). It can be a bit misleading, since in common sense, reader may think betting on vaccine means betting on vaccine can protect participants.

**Is the rationale for developing the new method (or application) clearly explained?**

Yes

**Is the description of the method technically sound?**

Partly

**Are sufficient details provided to allow replication of the method development and its use by others?**

Yes

**If any results are presented, are all the source data underlying the results available to ensure full reproducibility?**

No source data required

**Are the conclusions about the method and its performance adequately supported by the findings presented in the article?**

Partly

***Competing Interests:*** No competing interests were disclosed.

***Reviewer Expertise:*** DTA meta-analysis, Clinical prediction models, Health Technology Assessment

**I confirm that I have read this submission and believe that I have an appropriate level of expertise to confirm that it is of an acceptable scientific standard, however I have significant reservations, as outlined above.**

Author Response 03 Mar 2025

**Judith ter Schure**

We apologize for this very late reply. However late, we still hope to contribute to a complete and accurate scientific literature with a Version 2 of our paper that incorporates all the very useful comments made by the reviewers. Version 2 also further illustrates the unique features of the method by extending the example ALL-IN prospective meta-analysis in the paper with a reflection on the collaboration, summary of results and references to a complete replication package.

We thank Junfeng Wang for his thorough reading of our paper. We are especially grateful that he spotted a mistake in Section 1.1, pointed out as Minor comment 1, and was so kind to state "*I assume the authors forget adding something*". We indeed forgot to add the betting strategy, which we added in our Version 2 of the paper (see below at Minor comment 1). We also found a mistake ourselves in Section 1.3: there was a square root missing below equation (6), needing a division by 2 instead of 4 (which comes from the standard deviation of a Bernoulli-0.5 likelihood).

ALL-IN can indeed provide a "*different understanding of evidence synthesis, and a novel way of performing meta-analyses*", as stated by Junfeng Wang. Although ALL-IN is a novel statistical approach to meta-analysis, it does provides statistical summaries for evidence synthesis that can be reported in a familiar way – e-values take the role of p-values, and anytime-valid confidence intervals the role of conventional confidence intervals. Our worked-out example illustrates this in Version 2 of the paper. Yet due to their anytime-validity, this statistical summary can be *Live* and can be the *Leading* source of information in decisions on future research, even at *INterim* stages of clinical trials. So while the statistical approach fits within mainstream understanding of evidence synthesis as an external and retrospective summary, it can also shift that understanding towards evidence synthesis as an active/involved (advising future trials/liaising with ongoing trials) and possibly prospective tool to facilitate collaboration and increase efficiency.

In our Version 2 of the paper we have emphasized this aspect more, and moreover, stressed how the approach to prospective meta-analysis differs from other adaptive approaches to prospective analysis like the Framework for Prospective, Adaptive Meta-analysis (FAME), and Bayesian adaptive methods. The main difference is that the validity of the ALL-IN analysis does not rely on enforcing a maximum sample size or stopping rule, such that prospective collaborations can be more bottom-up rather than outsourcing decisions to an external top-down meta-analysis team. This fact needed a better illustration in Section 3 on Collaboration, so we rewrote that section by extending the discussion of our BCG-CORONA example with the results from that ALL-IN meta-analysis, thus including a completely worked-out example to the paper.

We will address all further comments in the order given. We hope that by proposing some adjustments to the paper in reply to each of his comments, we can convince Junfeng Wang that the description of the method is technically sound (question 2, now answered by 'Partly') and that the conclusions about the method and its performance are adequately supported by the findings presented in the article (question 5, now answered by 'Partly'). Hence we hope our Version 2 will be accepted without any further reservations.

Major comments

1. We agree with Junfeng Wang that our representation of the FDA game is approximate and not exact, simplifying the story such that we could calculate betting scores / *e*-values based on the COVID-19 event counts alone (to keep it simple as an illustrative example). Following Junfeng Wang's concerns, we will make this more explicit in Version 2 of the paper, after the brief mention halfway the first paragraph on the right side of page 3 (*'Most COVID-19 vaccine trials randomized large numbers of participants 50:50 vaccine:placebo such that we can assume that also throughout the trial the participants at risk stayed approximately balanced.'*). We add the following text and footnote there:

   *This allows for a back-of-the-envelope calculation to reinterpret the design for the COVID-19 vaccine trials in the language of betting\footnote{Note that in Section 1 (specifically at the end of Section 1.1) we discuss how to exactly analyze such time-to-event data using e-values taking into account that the risk set changes after occurrence of events and censoring. We expect such a reanalysis to be very well approximated by the simple calculations that are used for illustration in this introduction of the FDA game, at least for large and balanced trials like the one by Pfizer/BioNtech. Even for trials that are a lot smaller, this approximation is quite good. For example, the exact logrank e-value of 1,88 that we report for the NL trial in Table 1 can be approximated using this back-of-the-envelope calculation for the e-value as folllows:* $\left(\frac{80/180}{100/200}\right)^{96} \cdot \left(\frac{100/180}{100/200}\right)^{110} = 1{,}33$ *. In this trial only 1496 participants were 50:50 randomized (compared to the 40 thousand in the Pfizer/BioNtech trial), and 96 events observed in the treatment and 110 events in the control group (Ter Schure et al, 2022); testing a 20% VE against a 0% VE.}*

   We do believe that our approximation stays true to the essence of these large vaccine trials. For the Pfizer/BioNtech trial, Polack et al (2020) report a 95% vaccine efficacy based on a beta-binomial model taking into account the confirmed cases per 1000 person-years of follow-up based on 8 cases in the vaccine and 162 cases in the control group. This corresponds very well to a simple calculation (1 – 8/162)*100 = 95% assuming a perfect 50:50 ratio of follow-up time or indeed an infinite number of participants in each group.

The footnote points to existing text at the end of Section 1.1, the second paragraph of page 8 (in version 1 of the paper), where we do make a recommendation for actual use going forward. Just like a *p*-value logrank test, the *e*-value logrank test takes into account, as well-spotted by Junfeng Wang, that *'when the number of participants still at risk in one group decreases, the probability of the next event occurring in this group will also decrease'*. Moreover, the exact *e*-value logrank test can be used in sequential settings in which the *p*-value logrank test fails, such as in sequential analysis in unbalanced randomized experiments (where p-value tests cannot be used based on group-sequential or alpha spending methods that rely on Gaussian approximations). As we write in the paper: *We will use the logrank Z-statistic as a running example for meta-analysis on summary statistics. For an IPD meta-analysis (on individual patient data), however, we recommend to use the exact e-value logrank test from Ter Schure et al. (2024) that is valid regardless of the randomization (e.g. 1:1 balanced or 1:2 unbalanced), the number of participants at risk, the number of events or the size of the effect.*

2. We believe we do address heterogeneity when we state the following on page 14: *'heterogeneity in their effect sizes (e.g. one 20%, one 50%, one 25%) does not matter for their joint ability to reject the global null hypothesis of no effect in all trials. So for testing the global null, trials are allowed to be heterogeneous in where they are in the space of the alternative hypothesis H1 = {VE: 20%* ≤ VE ≤ 100%}. For estimation, however, it is not clear what the ALL-IN confidence interval is estimating if we assume that the effects in the trials are very different. Still, as a first summary, a typical effect size (Peto, 1987) might be useful if we are unable to estimate a random effects model. The development of confidence sequences for random-effects meta-analysis is a major goal for future work. We do not, however, believe that the evidence in a line of research should be monitored based on whether this interval excludes the null hypothesis, or whether the e-value corresponding to the random-effects null model does: for testing, the global null is much more natural. Waiting for a random-effect model to reach a certain threshold is counter-intuitive, since it might require many small trials to estimate the between-trial variability instead of focusing on testing the treatment effect. Moreover, the goal of rejecting the null hypothesis corresponding to this model can be quite strange. When testing a zero-effect null hypothesis, it assumes that there are true effects of harm and true effects of *benefit among the trials and that their mean is exactly zero.'*

3. Junfeng Wang is right to point out that the COVID-19 vaccine trials are a very special example showing how meta-analysis can be done on interim trial results when many studies are simultaneously ongoing. As also stated in the abstract, ALL-IN meta-analysis can be performed on interim data, but does not have to. The analysis design requires no information about the number of patients in trials or the number of trials eventually included. In Section 1.2 we first give the example where we can collect a Z-statistic for each study out of K studies completed so far, so in a setting of conventional retrospective meta-analysis. What makes ALL-IN meta-analysis special is stated right after: the analysis on interim results is exactly the same. The connection to living systematic reviews is made earlier in the paper, on page 5 where we refer to Simmonds et al (2017). Conventional meta-analysis does not control type-I error rates in living systematic reviews, while ALL-IN meta-analysis does. We do see the need to make this connection more explicit: allowing updates of systematic reviews based on interim data is what "breathes life into living systematic reviews". Also following another reviewer's remarks, we will use our example of ALL-IN-META-BCG-CORONA in Section 3 to make more explicit what ALL-IN methods can do that is not possible with any other approach.

4. Future research should definitely focus on whether the method can be extended to network meta-analysis, but we have not conducted such research yet. We are pleased to see that the paper convinced the reviewer that the method is easily extended to IPD meta-analysis. We have added these considerations to the discussion section of the Version 2 paper.

5. Covariate adjustment is a major field of research in the literature on *e*-values and anytime-valid statistics. We propose to add the following paragraph to the discussion section in Version 2 of the paper:

   *This paper introduces ALL-IN meta-analysis based on Z-score methods of meta-analysis, as introduced in standard works like by Borenstein et al. (2009). We mainly focus on testing rather than estimation. For testing, extensions to IPD meta-analysis based on exact e-values (rather than Z-score approximations) follow easily since both can be combined by multiplication, as shown in our ALL-IN-META-BCG-CORONA example. Within this focus on testing with type-I error control, the heterogeneity question is less explicit, since under the null-hypothesis there is no heterogeneity, as discussed in Section 3.3. In future work we will provide more details on estimation under heterogeneity, with anytime-valid confidence intervals for the fixed-effect (singular), fixed-effects (plural) and random-effects model. Extensions to network meta-analysis seem possible, but are not yet our main direction. With regard to meta-regression, much development can be expected from work on e-values in settings of composite null-hypotheses, for which research is ongoing that extends to linear regression (e.g. Pérez-Ortiz et al., (2024)). For IPD meta-analyses that take covariates into account, analysis of randomized controlled trials can use the so-called 'Model-X' e-value approach.*

Pérez-Ortiz, M. F., Lardy, T., de Heide, R., & Grünwald, P. D. (2024). E-statistics, group invariance and anytime-valid testing. The Annals of Statistics, 52(4), 1410-1432.

Grünwald, P., Henzi, A., & Lardy, T. (2024). Anytime-valid tests of conditional independence under model-X. Journal of the American Statistical Association, 119(546), 1554-1565.

Minor comments

1. Junfeng is completely right that we forgot something: the betting strategy. We

propose to correct the text as follows:

*For example putting 1/3 on vaccine and 2/3 on placebo: 1/3· 70/170 · 170/70 + 2/3· 100/170 ·170/100 = 1. No matter how we invest in the two outcomes, (e.g. try putting 1/2 on vaccine and 1/2 on placebo, or something different) in expectation under the null we multiply the initial investment by 1.*

2. This is a great comment! Indeed the ones betting on development of an effective vaccine – e.g. the NIH in funding the trials (taking 'multiple shots on goal' to just 'let the chips fall') – hope to find very few events of COVID-19 in the vaccine group of a clinical trial. We will adjust how we phrase this to eliminate this confusion in Version 2 of the paper.

***Competing Interests:*** No competing interests were disclosed.